\documentclass{article}

\usepackage{arxiv}

\usepackage[utf8]{inputenc} 
\usepackage[T1]{fontenc}    
\usepackage{hyperref}       
\usepackage{url}            
\usepackage{booktabs}       
\usepackage{amsfonts}       
\usepackage{nicefrac}       
\usepackage{microtype}      
\usepackage{lipsum}
\usepackage{graphicx}
\usepackage{float}
\graphicspath{ {./images/} }

\title{Doubly Forced Anharmonic Oscillator Model for Floating Potential Fluctuations in DC Glow Discharge Plasma}

\author{
 K. Jayaprakash$^{1}$, Prince Alex$^{1,2}$, A. Saravanan$^{1,3}$, M. Perumal$^{1}$, Thangjam Rishikanta Singh$^{1}$, 
 and Suraj Kumar Sinha $^{1 \ast}$\\\\
  $^{1}$ Department of physics, 
 Pondicherry University,
  India, 605014 \\

  $^{2}$ Departimento di Fisica G.Occhialini, Universitadeglistudi di Milano-Bicocca, 20126, Italy. \\

  $^{3}$Institute for plasma research, Gandhi Nagar, Gujarat,382428, India.\\\\
  $^{\ast}$\textbf{\texttt{sinhasuraj.phy@pondiuni.edu.in}}
  
}

\begin{document}
\maketitle
\begin{abstract}
The Floating Potential Fluctuations (FPF) observed in a dc glow discharge plasma powered with two sources is modeled using an anharmonic oscillator with two forcing terms. In the discharge system, one of the electrode is biased to a negative voltage source (i.e. cathode), and the second electrode is biased to a positive voltage source (i.e. anode), while the stainless-steel vacuum chamber is grounded. The dc glow discharge plasma is generated by application of negative voltage on the cathode with respect to the grounded chamber using one of the power supplies. On application of positive voltage to the anode using second power supply results in formation of potential structure on achieving the triggering criteria. This potential structure is referred as anodic double layer (ADL). The evolution of ADL is associated with FPF. Therefore, FPF is analyzed to characterize the ADL's dynamical features. In this work, the experimentally observed FPF compared with numerically obtained oscillations using an anharmonic oscillator model with two forcing terms. Each of these forcing terms are associated with the two power supplies used in the experiment. The experimentally and numerically obtained oscillations from the model are studied using phase-space plot, FFT, Largest Lyapunov exponent (LLE). The dynamical features of oscillations obtained by the model show strong agreement with the experiment and can be extended for a description of complex systems driven by multiple forces.
\end{abstract}


\section{Introduction}
The sheath dynamics are of prime importance for understanding the governing mechanism of a discharge. Accordingly, sheaths have been investigated since the beginning of  $20^{th}$ century \cite{ref4.1}.  In a dc glow discharge system, sheaths are the junctions between plasma and electrodes confining it \cite{ref4.2}. Generally, two plasma-metal junctions are formed, corresponding to cathode and anode, for such plasmas. The ions shrouding cathode, form ion-sheath (also known as cathode sheath), and the electrons shielding anode form anode-sheath (also known as electron sheath) \cite{Lieberman}. The sheaths, act to balance ion and electron flow to maintain the discharge in the system\cite{ref4.4}. The ion sheaths are widely investigated because of its primary role in processing plasmas \cite{ref4.4, ref4.5, ref4.6}. However, electron sheaths has been in focus for two decades \cite{ref4.7, ref4.8, ref4.9, ref4.10, ref4.11, ref4.12, ref4.13, ref4.14, ref4.15, ref4.16, ref4.17, ref4.18, ref4.19, ref4.20, ref4.21, ref4.22, ref4.23, ref4.24, ref4.25, ref4.26}. In laboratory plasmas, electron sheath is a thin region near the anode where the electron density increases monotonically from plasma to the anode surface \cite{ref4.5}. Several new features of the electron sheaths have been observed  recently, and ADL is most interesting among them \cite{ref4.27,ref4.28}. Double layers (DL) are the source of various dynamical features of charged fluids in biological systems \cite{ref4.29,ref4.30}, solid-state batteries \cite{ref4.31}, and space and astronomical plasmas \cite{ref4.32}. Furthermore, Nobel laureate H. Alfen advocated that many unknown cosmology features can be understood by scaling up DL laboratory observations \cite{ref4.34}.
	
Recently, we have studied dynamical features of ADLs including order-chaos-order-chaos transitions via a periodic doubling route\cite{ref4.27}, boundary conditions for the formation of DLs \cite{ref4.35}, self organized criticality, coexistence of chaos and complexity dynamics under typical ADL condition \cite{ref4.28,alex2018self}. In these investigations, stable ADL generated in a dc glow discharge plasma system, where, each electrode (i.e. cathode and anode) is powered by separate sources with respect to grounded chamber. The FPF were analyzed under different operating conditions to understand the evolution of ADL’s dynamical features \cite{ref4.27}. For this experimental system, a need of a model  has arisen to explain the governing dynamics. A doubly forced anharmonic oscillator (DFAO) model fits this requirement to describe dynamical features of FPF observed in the discharge system.

The harmonic oscillator is a standard model and widely applied to describe many physical systems \cite{ref4.36}. Plasma is a non-linear system and its dynamics has been studied in the past using Van der Pol oscillator\cite{ref4.37,ref4.38,ref4.39,ref4.40,ref4.41,gyergyek1999experimental}, Duffing oscillator \cite{ref4.42}, and anharmonic oscillator\cite{ref4.43,ref4.44,ref4.45}. Charney, and Hasegawa-Mima developed a non-linear model for the context of geophysical weather and magnetized plasma, respectively \cite{ref4.36}. In plasma based experiments different dynamical features of FPF has been reported such as  period oscillation to chaos via periodic-doubling route \cite{ref4.46,ref4.47}, quasi-periodic route \cite{ref4.48,ref4.49}, intermittent route \cite{ref4.50}, period-adding and period pulling routes \cite{ref4.52} and period-subtracting routes in different systems \cite{ref4.51}. Further, various interesting transitions between ordered and chaotic dynamical behavior has been observed experimentally \cite{ref4.53,ref4.54} for example, order-chaos-order-chaos transition \cite{ref4.27,ref4.55}, irregular to regular mixed modes transitions \cite{ref4.56}, mode-suppression \cite{ref4.57}, homoclinic \cite{ref4.58}, monoclinic bifurcation \cite{ref4.59,ref4.60}, self-exited \cite{ref4.61} and relaxation oscillation \cite{ref4.62}. To study dynamical features of nonlinear systems time series analysis is important and widely applied for weather forecasting \cite{ref4.63,ref4.64}, ecological study\cite{ref4.65}, cardiac and neural tissue electrodynamics \cite{ref4.66,ref4.67,ref4.68}, turbulence in magnetic fusion machines \cite{ref4.71,ref4.72,ref4.73}. 

Mahaffey in 1976 \cite{ref4.42} presented unforced anharmonic oscillator model for plasma oscillations having features of shift frequency of oscillation and asymmetry of amplitude of oscillations and changes in the resonance response curve. Kadji et. al.\cite{ref4.43} modified this model for forced anharmonic oscillator and a similar model was used to describe FPF in a dc glow discharge plasma system with constricted anode geometry \cite {ref4.45}. In this work, we introduce two forcing terms in the anharmonic oscillator model to demonstrate FPF in the dc glow discharge, powered by two sources. The order-chaos transitions of FPF observed in the experiment are regenerated numerically using the DFAO.	

In the next section, the detailed experimental arrangements are presented. Section 3 gives details of the anharmonic oscillator model and the proposed DFAO model. Followed by an analysis of FPF observed in the experiment and generation of similar oscillations numerically using the model. The dynamical behavior of these oscillation characterized by phase space plot, FFT, Largest Lyapunov exponent, and phase space reconstruction is discussed in section 4. Followed by discussion and conclusions.  
\section{Experimental set-up }
\begin{figure*}
	\centering
	\includegraphics[width=0.9\textwidth]{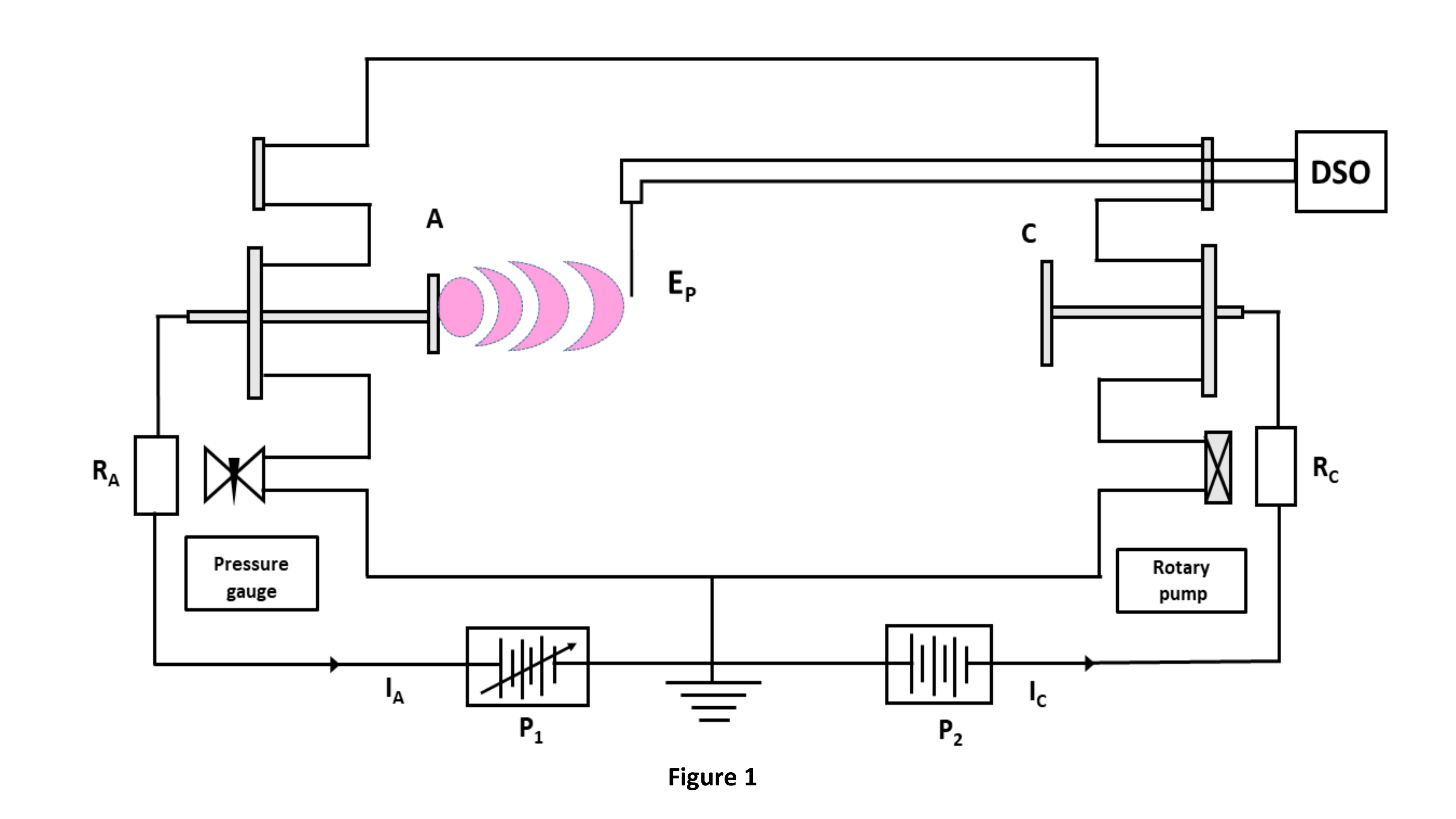}
	\caption{Schematic representation of the experimental setup, $E_{P}$-Electrostatic Probe, C - Cathode, A- Anode,  $R_{A}$ -Anode Resistance, $R_{C}$-Cathode Resistance, $P_{1}$- Anode Voltage,$P_{2}$- Cathode Voltage $I_{A}$-Anode Current,  $I_{C}$– Cathode Current}
	\label{fig:exprimental setup}      
\end{figure*}

The schematic diagram of the experimental setup is shown in Fig.(\ref{fig:exprimental setup}). 
The setup consists of an SS304 cylindrical chamber of length 730 mm, diameter 300 mm, with four glass windows for visual observations. A disc shaped stainless-steel (SS) cathode with a diameter of 50 mm, thickness 5 mm, is negatively biased with respect to grounded chamber to a high voltage dc power supply $P_{1} $(1 kV, 1 A). The anode is a tungsten (W) disc with 20 mm diameter and 2 mm thick, biased positively using another dc power supply $P_{2}$ (750 V, 1 A). The current limiting resistors, $R_{C}$ and $R_{A}$ $(100  \Omega )$ were connected across the cathode and anode power supplies respectively. The current $I_{C}$ and $I_{A}$ were measured across the resistors $R_{C}$ and $R_{A}$. The experiments were carried out at a pressure of 0.3 mbar by using air as the background gas \cite{ref4.28}. Glow discharge plasma was first generated between the cathode and grounded chamber. 

The cathode voltage is fixed at 450 V. Followed by application of positive bias voltage on the anode in range of 0 - 200 V, which resulted in the streaming of plasma electrons towards the anode. Anode voltage increased slowly, and fluctuation are observed. At a threshold value of anode bias, i.e., at 40 V, a visible ADL structure formed, it evolved into multiple anodic double layer (MADL) and gradually decayed with monotonic increase of anode voltage. An electrostatic probe $(E_{P})$ is placed along the cylindrical chamber's axis to record the FPF at different stages of evolution of the ADL. The time series of the FPF were recorded with a high band Digital Storage Oscilloscope (LeCroy Wavejet-354A, 500 MHz). In next section the DFAO model for this experimental arrangement is given in detail. 
\section{Doubly forced anharmonic oscillator (DFAO) model}
Like other physical systems, the nonlinear oscillations emerging from the density fluctuations in plasma
under internal or external perturbation can be modeled as a nonlinear harmonic oscillator. Various harmonic
oscillator models such as Vander Pol oscillator\cite{ref4.37,ref4.38,ref4.39,ref4.40} duffing oscillator \cite{ref4.41} and anharmonic oscillator\cite{ref4.42,ref4.43,ref4.44}  are
 employed to describe plasma oscillations under different conditions. In 1976, the first such attempt was
made by Mahaffey. The two-fluid equations that govern the dynamics of ion and electrons in the plasma,
reduced to the form of an anharmonic oscillator model and it gives the evolution of plasma density
 fluctuation as
\begin{equation}
	\ddot{x}+\omega ^{2}y+\alpha y^{3}+d=0
	\label{eqn:1}
\end{equation}
Here, the over dot on the variable denotes time derivative, x-denotes plasma potential or density fluctuation
 and $\omega$ is the natural frequency of oscillations, $\alpha$ is the cubic nonlinear parameter, and d is the constant external force applied. In another work, the Duffing oscillator model proposed by Mahafey was modified
by Kadji et al. to demonstrate the regular and chaotic behaviors of plasma oscillations \cite{ref4.43,ref4.44} as 
\begin{equation}
	\ddot{x}+\omega ^{2}x+\beta x^{2}+\alpha x^{3}+d=E cos \Omega t
	\label{eqn:2}
\end{equation}

Here, $\beta$ is the is a quadratic nonlinear parameter, and $E cos \Omega t$ is the external sinusoidal driving force acting
on equilibrium plasma with and amplitude of E and range of driving frequency of $\Omega$. This forced nonlinear 
harmonic oscillator model was further improved to mimic the exact nature of plasma oscillation under DL
condition obtained in constricted anode geometry as \cite{ref4.45}
\begin{equation}
	\ddot{x}+X_{1}\dot{x}+x
	+M X_{1}
	 =A f\left( 1+f cos \omega_{1} t\right)  
	\label{eqn:3}
\end{equation}
where, $X_{1}$$\,=\,$$\left( M+P+Nx+Ox^{2}\right)$,
\,\,\,\,\,$X_{2}$$\,=\,$$\left(Px+\frac{N}{2}x^{2}+\frac{O}{3} x^{3} \right)$, $\omega_{0}=kC_{s}$ $k$ is the wave number, $C_{s}$ is ion acoustic speed. The other parameters such as $M=\frac{\nu_{i}}{\omega_{0}}$, $N=\frac{\alpha}{\omega_{0}}$, $O=\frac{\mu}{\omega_{0}}$, and $P=\frac{\lambda}{\omega_{0}}$ respectively denote collisional, ionization, two body and three-body recombination frequencies normalized by the natural frequency of ion-acoustic oscillations. The ion-acoustic oscillation frequency can be written as
\begin{equation}
\omega_{0}=k\left(\frac{k_{B}T_{e}}{m_{i}}+\frac{k_{B}T_{i}}{m_{i}}\right)\label{eqn:4}
\end{equation}
Where $k-$ is the wavenumber and $m_{i}$ is ion mass, $k_{B}$ is Boltzmann constant, and $T_{e}$ is plasma electron
temperature and $T_{i}$ is plasma ion temperature. Since, $T_{e} \gg T_{i}$ in low-temperature non-equilibrium plasma,
one can neglect the contribution of ion velocity term. For a range of discharge conditions, this ion-acoustic
 frequency can be estimated provided the value of electron temperature is known. The natural frequency of
 oscillation, $\omega_{0}$ is obtained by solving Eq.(\ref{eqn:4}), assuming the parameters $M=N=O=P \approx 0$ The term A, on the right-hand side is added to account the effect of discharge voltage in addition to the plasma generated
  normal mode oscillations
   \begin{table*}
	\centering
	\caption{Range of anode voltage (V) and the force factor $B$ used with the nature of numerical and experimentally oscillations corresponding to the cathode voltage of $-450 V.$}
	\begin{tabular}{lllll}
		\hline\noalign{\smallskip}
		Anode &  Figure &Amplitude &  Figure  &Nature  \\
		Voltage (V)& Number&Forcing (B)& Number  & oscillation\\
		& (Experiment) && (Numerical)  & \\
		\noalign{\smallskip}\hline\noalign{\smallskip}
		81	&2\,\textbf{(a)}	&0.01	&2\,\textbf{(g)}	&Periodic Oscillation \\
		93	&2\,\textbf{(b)}	&0.5	&2\,\textbf{(h)}	&Period Doubling\\
		97	&2\,\textbf{(c)}	&0.65	
		&2\,\textbf{(i)}	&Chaotic nature	\\
		109	&2\,\textbf{(d)}	&1		&2\,\textbf{(j)}	&Periodic nature\\
		112	&2\,\textbf{(e)}	&2		&2\,\textbf{(k)}	&period doubling \\
		160	&2\,\textbf{(f)}	&2.7	&2\,\textbf{(l)}	&Chaotic nature\\
		\noalign{\smallskip}\hline
	\end{tabular}
	\label{Table}
\end{table*}
Many physical mechanisms could cause global collective plasma oscillations. Few such mechanisms are
external oscillating electric field, the passage of laser into the equilibrium plasma, beam of electrons or ions
driven into the plasma, passage of dust into the plasma medium, density gradient, and pressure gradient
etc . As discussed in the experiment, we model our plasma oscillations produced under MADL conditions obtained using two power sources. In this experiment, one power supply
was used to bias the cathode to sustain the background plasma between the cathode and chamber, and the
power supply was used to power the anode with respect to the common grounded chamber. Variations in
the voltages of these power supply change the conditions of plasma between anode and cathode, which has
significant effects on the transport of ions and electrons in the plasma, which ultimately forms a series of DL. The formation of multiple DL between anode and cathode triggers different modes of nonlinear plasma
oscillations for a range of anode and cathode voltages. Therefore, to model these plasma oscillations for the present experimental conditions, which supports the formation of multiple DLs for different discharge conditions, a new forcing term has been added on the right-hand side of Eq. (\ref{eqn:3}) to mimic the force introduced
by the second power source. A detailed derivation of this expression. In Eq.(\ref{eqn:3}), x-denotes the floating potential 
\begin{equation}
\ddot{x}+X_{1}\dot{x}+x
+M X_{1}
=A {f_{1}\left( \omega_{1} \right)} +Bf_{2}\left(\omega_{2} \right)   
\label{eqn:5}
\end{equation}
Where,  $f_{1}(\omega_{1})=(1+f_{1}cos \omega_{1} t)$, $f_{2}(\omega_{2})=(1+f_{2}cos \omega_{2} t)$, $B$ is the force constant, $f_{1}$ and $f_{2}$ are the amplitude and frequency of forces representing the power supplies. Since the application of a second power source adds extra force to the plasma species, it is represented by a positive sign. In other words, the application of positive voltage on the anode is equivalent to increasing discharge voltage. The addition of factor $B$ accounts for the contribution of regular mode oscillation produced by the second source. Eq.(\ref{eqn:5}) can be solved numerically with the appropriate nonlinear coefficients. The values $ M, N, O, P \approx 0.3$, the values of $f_{1}$ and $f_{2}$ are taken as 0.36, -1.6 respectively, these values are arrived by iteration method.
 The $M, N, O, P,$ $f_{1}$, and $f_{2}$ are kept constant throughout the numerical calculations. The value of $B$ is varied, as shown in table.(\ref{Table}). numerically regenerate FPF to mimic precisely the nature of experimentally obtained
 fluctuations for anode voltages
 parameters for which the numerically obtained fluctuations matched the experimental observations are
 listed in Table.\ref{Table},  numerically regenerate FPF to mimic precisely the nature of experimentally obtained
 fluctuations for anode voltages.
\section{Analysis of FPF}
As mentioned above, we have generated typical glow discharge plasma by applying a potential between
 the cathode and grounded chamber. Once the discharge strikes at minimum breakdown voltage, we further
 increased the cathode to a reasonably high value of $- 450$ V to ensure stable, steady-state plasma, denoted
 using $A$ in the numerical model. ADL is generated in front of the anode by raising its potential to the
 triggering condition i.e, $40$ V. Once the stable ADL is generated, its further evolution was controlled by
varying anode bias between 81 V and 160 V. This is denoted using $`B'$ in the numerical model. Here in this
 section, we quantitatively compared the experimentally observed transition in FPF associated with ADL to  that generated using the numerical model described above to test the validity of our model. Different nonlinear techniques such as phase space trajectories, Fast Fourier Transform (FFT), and largest Lyapunov exponent were employed to quantify the dynamics of fluctuations along with visual observation of generated oscillations.
 \section{Oscillations}
\begin{figure*}[h]
	\centering
	\small{Experiment\,\,\,\,\,\,\,\,\,\,\,\,\,\,\,\,\,\,\,\,\,\,\,\,\,\,\,\,\,\,\,\,\,\,\,\,\,\,\,\,\,\,\,\,\,\,\,Numerical}\\
	\includegraphics[width=0.28\textwidth]{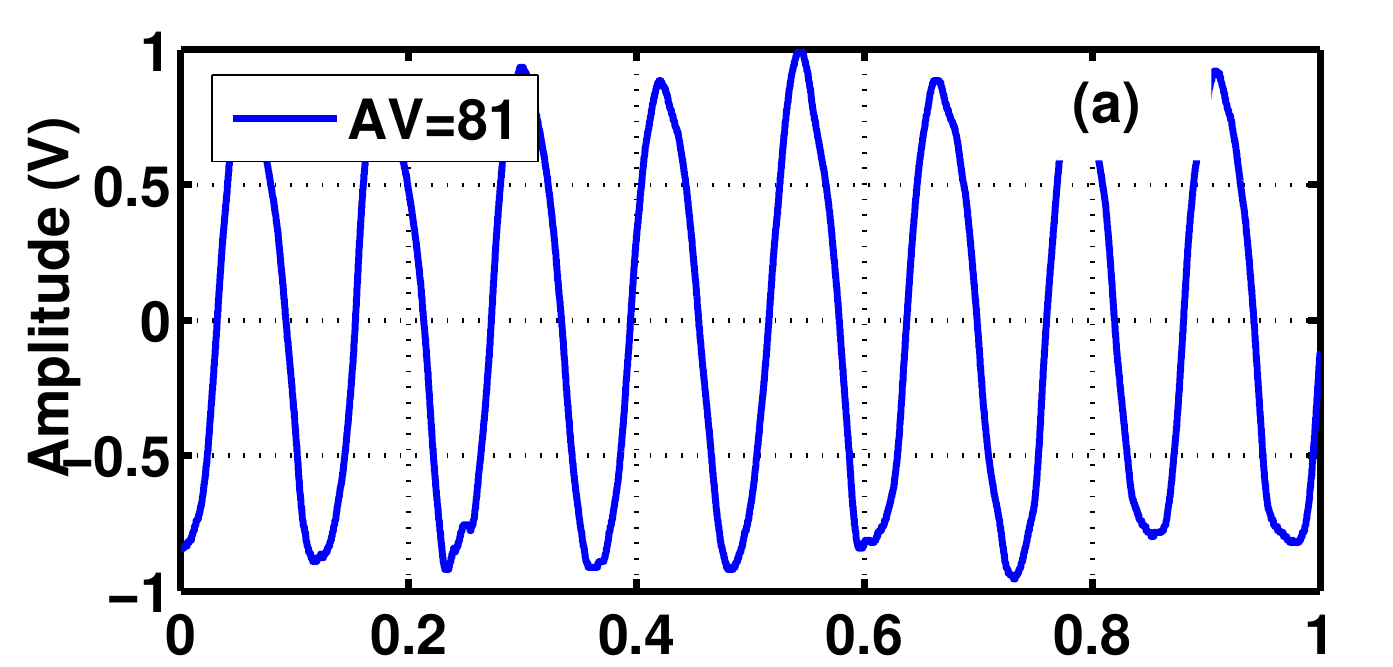}
	\includegraphics[width=0.28\textwidth]{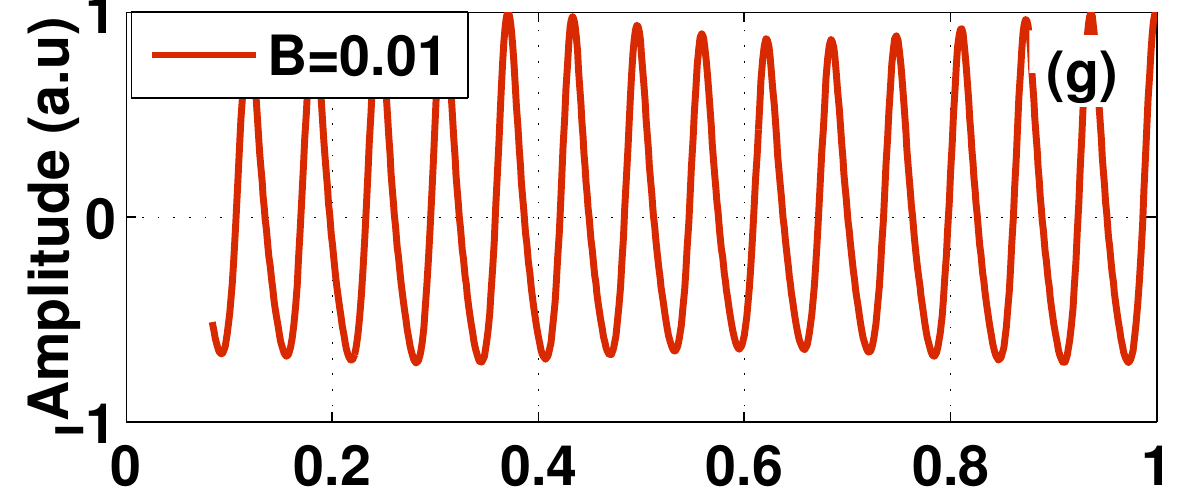}\\
	\includegraphics[width=0.28\textwidth]{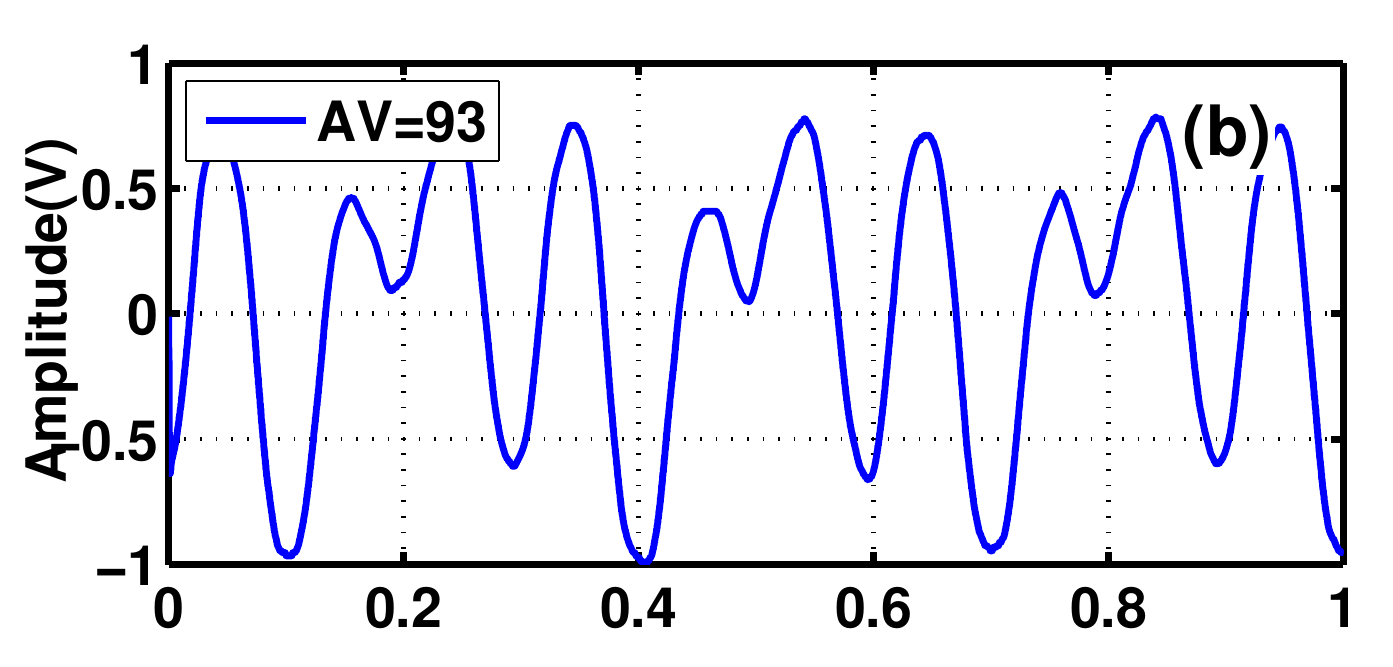}
	\includegraphics[width=0.28\textwidth]{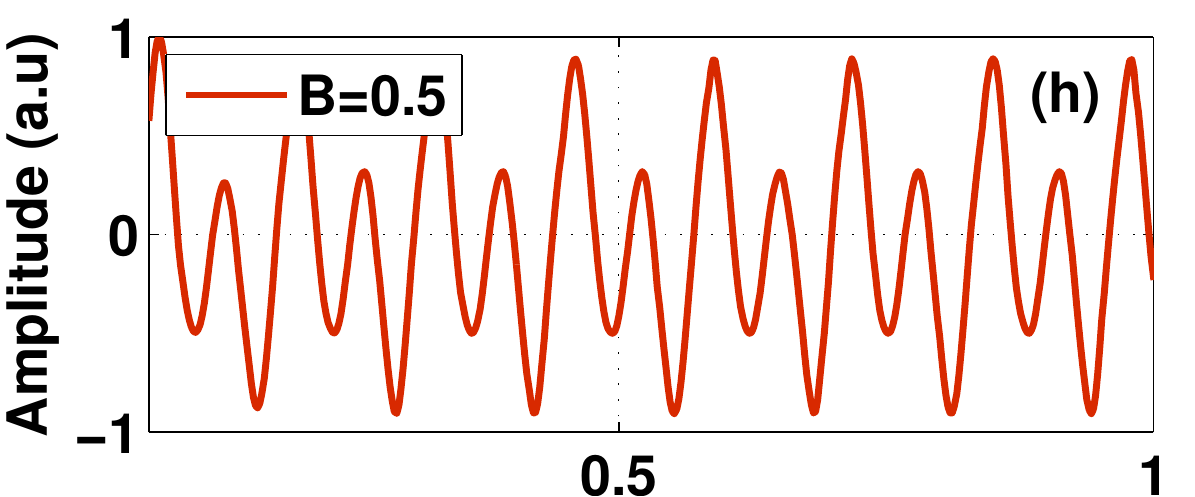}\\
	\includegraphics[width=0.28\textwidth]{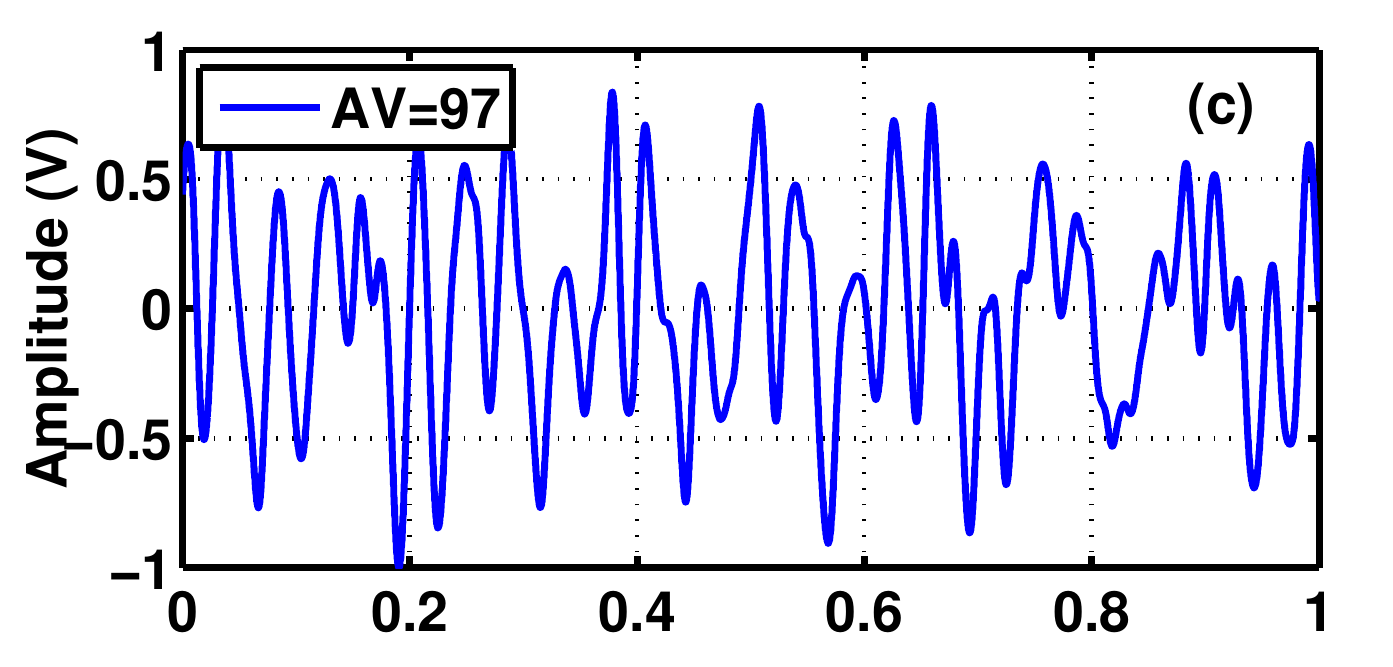}
	\includegraphics[width=0.28\textwidth]{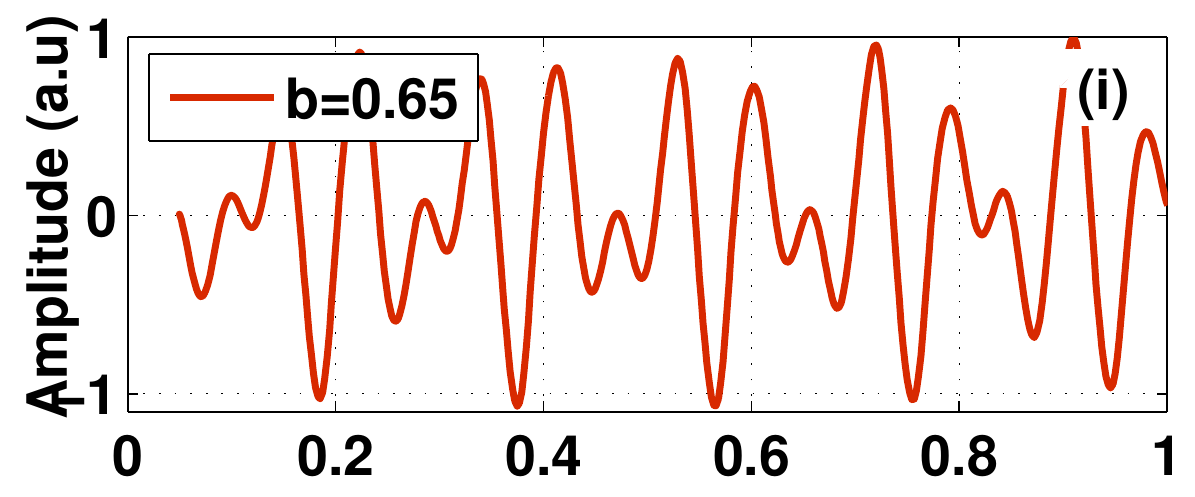}\\
	\includegraphics[width=0.28\textwidth]{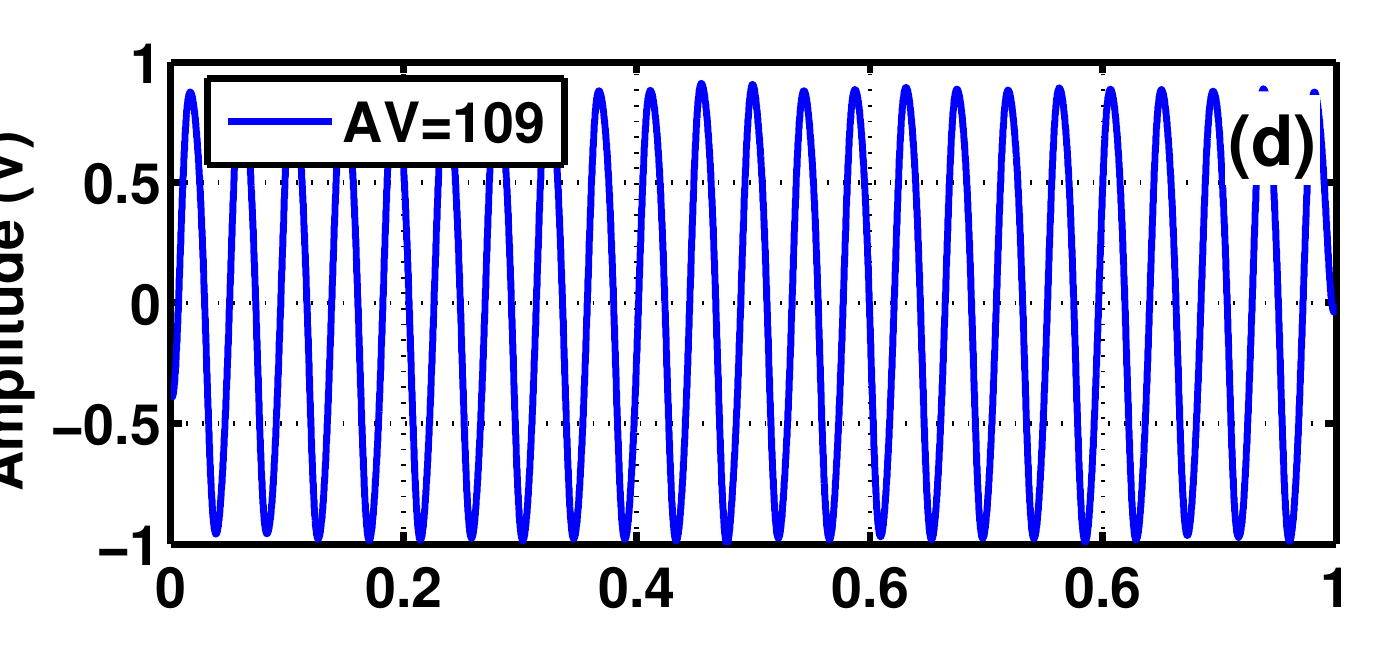}
	\includegraphics[width=0.28\textwidth]{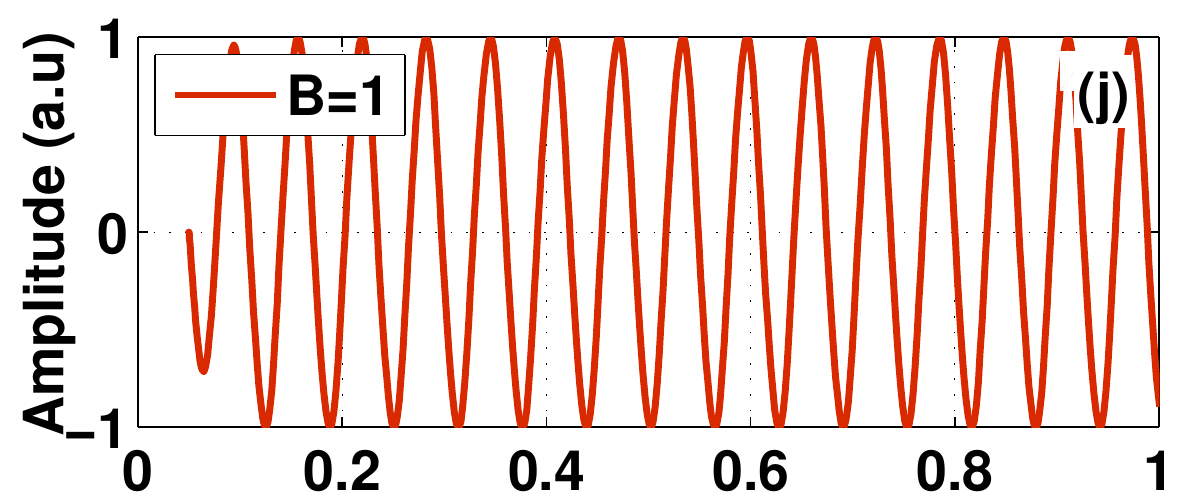}\\
	\includegraphics[width=0.28\textwidth]{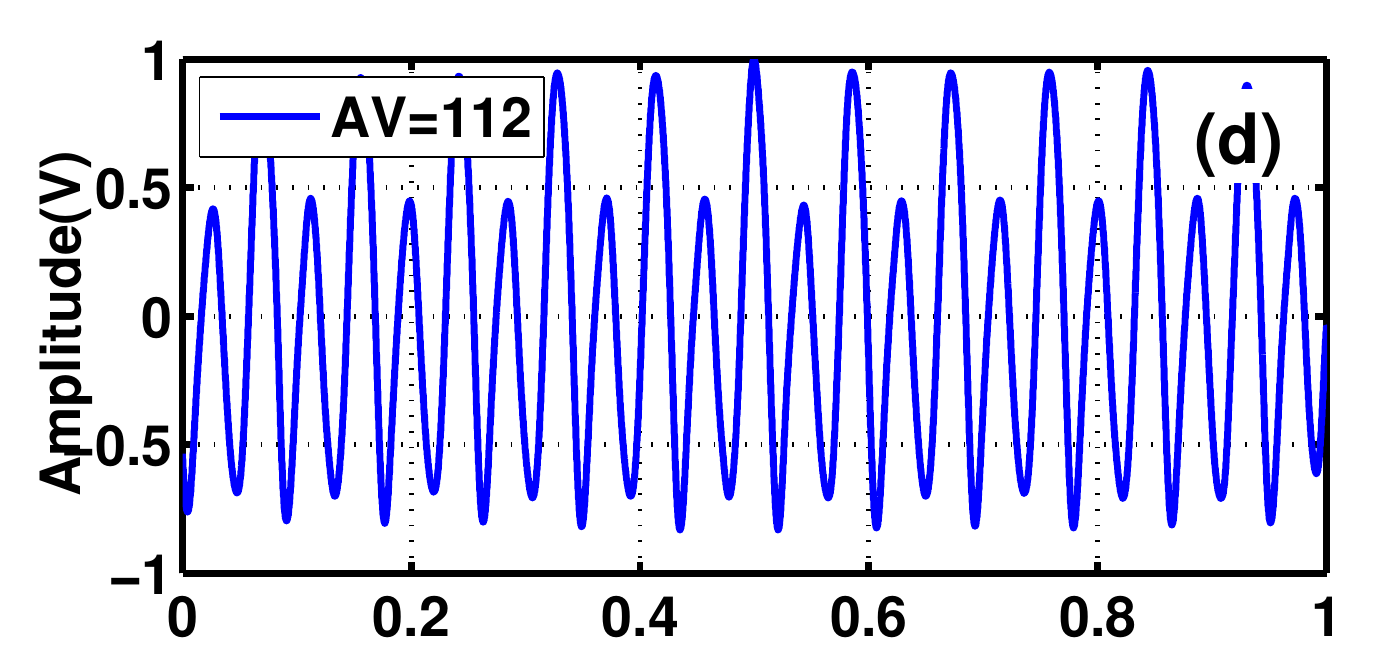}
	\includegraphics[width=0.28\textwidth]{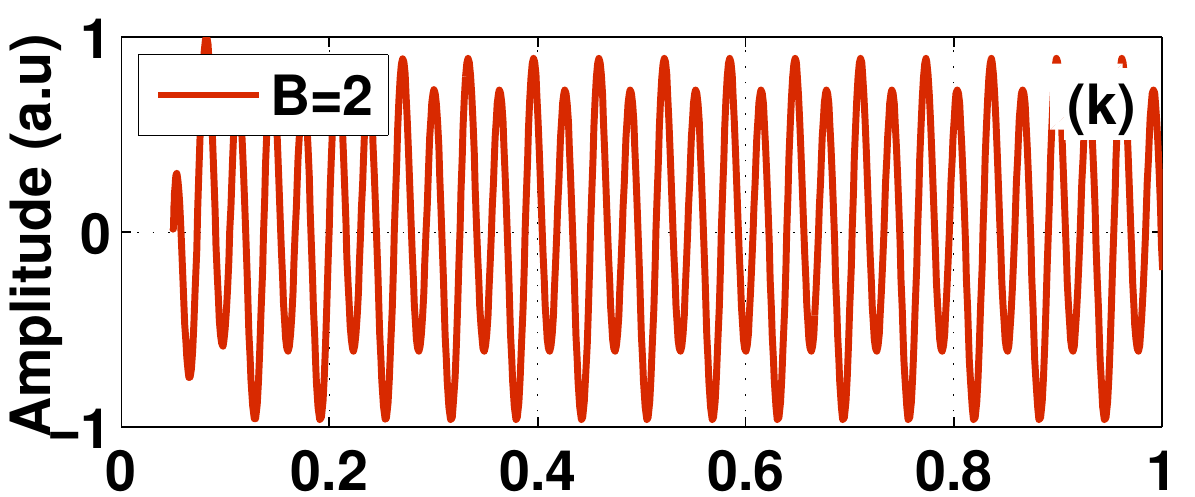}\\
	\includegraphics[width=0.28\textwidth]{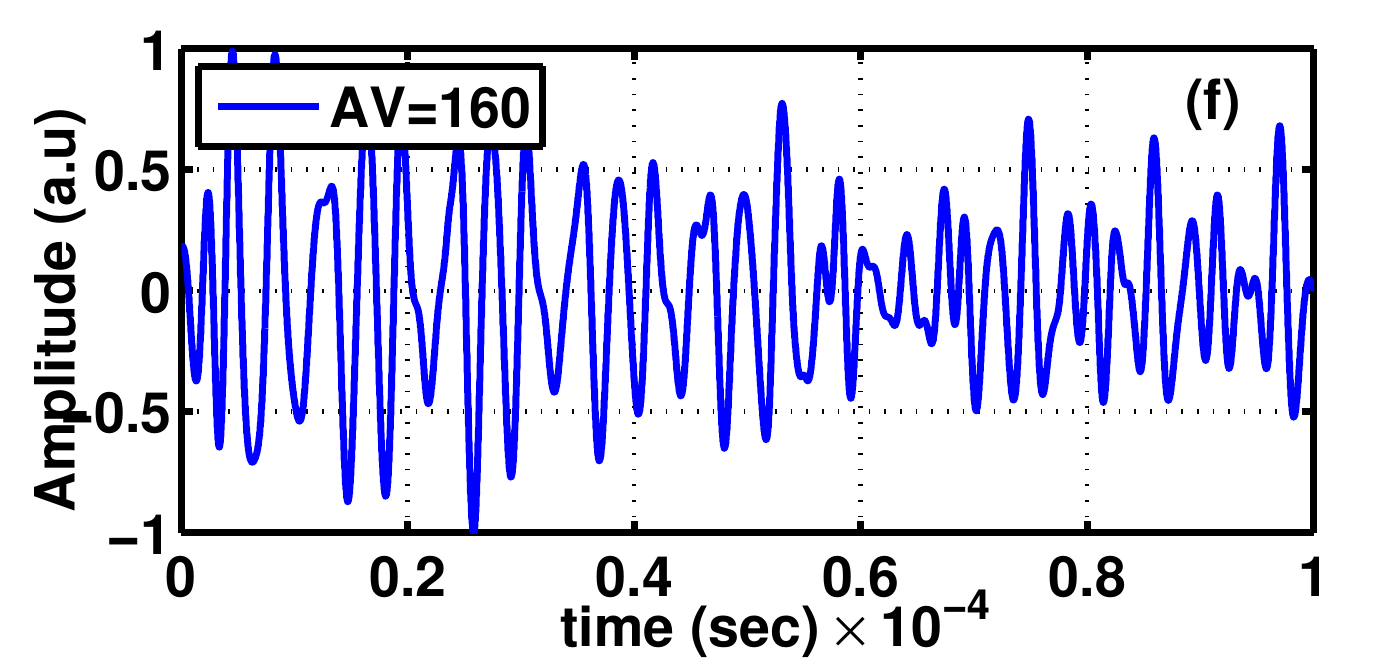}
	\includegraphics[width=0.28\textwidth]{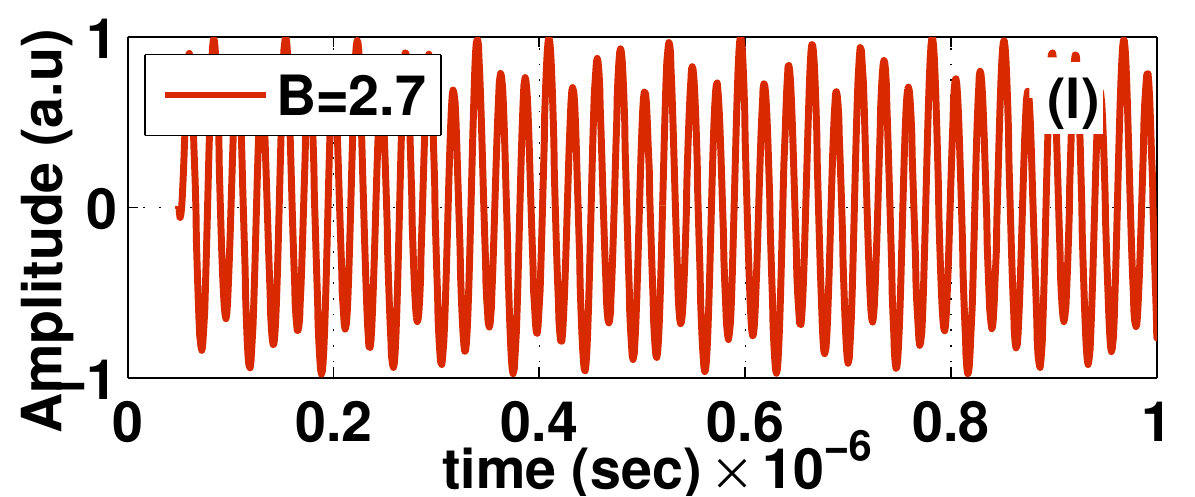}
	\caption{In Fig. 2. the first column   plots of 2(a)-2(f) are depicted  FPF obtained by the experimentally at different AV's are:(a)81 V, (b)93 V, (c)97 V, (a)109 V, (a)112 V, (a)160 V. The second column is indicating the FPF, generated  numerically by Eq. (\ref{eqn:5}), at different amplitude forcing parameter value B's are 2(g)-2(l):(g) 0.01, (g) 0.5, (g) 0.65, (g) 1, (g) 2, (g) 2.7  }
	\label{fig:2 oscillation}       
\end{figure*}

Experimentally observed as well as numerically generated fluctuations are shown in the Fig.\ref{fig:2 oscillation}, left (a-f) and right (g-l) panels respectively. An electrostatic probe fixed at 30 mm axially away from the anode
electrode is used to obtain experimental fluctuations. As shown in figure, fluctuations in Fig.\ref{fig:2 oscillation} (a) to Fig.\ref{fig:2 oscillation}  (f) corresponding to different states of ADL obtained at different anode voltage (AV) of 81 V, 93 V, 97 V, 109
V, 112 V and 160 V. Similarly figure 2 (g) to (l) generated for different values of $B$= 0.01, 0.5, 0.65, 1, 2 and 2.7 respectively. At AV=81 V, as shown in Fig.\ref{fig:2 oscillation} (a), the experimental observations show nearly periodic oscillations with period-1 behavior. This fluctuation corresponds to the initial stable state of ADL, representing a highly
 ordered state of the DL. These oscillations were modeled with $B=0.01$ using Eq.\ref{eqn:5} Upon
 increasing the anode voltage to AV =93 V in steps of 1 V, we found that two alternate dark and bright
 plasma regions form in front of the anode. These alternate dark and bright regions with visually clear
 boundaries constitute 3 ADLs, representing localized positive and negative space charges. The physical
aspects of a similar transition are described elsewhere \cite{ref4.27,ref4.61}. Here the system undergoes a period-doubling
 from a period-1 oscillation. Period- 2 limit cycle motion in the phase space shows the period-doubling
 nature. The ADLs begin to contract towards the anode with an increase in the discharge voltage, and at
 AV=97 V and corresponding phase space plot confirm chaotic nature shown in Fig.\ref{fig:2 oscillation}(c) and Fig.\ref{fig:3 Phase space}(c). Similarly,
 further increasing the anode voltage bias
 as follows AV =109 V, AV =112 V, AV =160 V, and forcing
 parameter $B$ values are $B=1$, $B=2$, $B=2.7$. The oscillation undergoes again periodic, period-doubling nature
 and then chaotic nature, its indicate the Fig.\ref{fig:2 oscillation} (d), 2(e), 2(f) and 2(j), 2(k), 2(l). We have collected too many oscillations; however, these oscillations were away from the model and experimental fluctuation.
 \section{Phase-space}
\begin{figure*}[h]
	\centering
	\small{Experiment\,\,\,\,\,\,\,\,\,\,\,\,\,\,\,\,\,\,\,\,\,\,\,\,\,\,\,\,\,\,\,\,\,\,\,\,\,\,\,\,\,\,\,\,\,\,\,\,\,\,\,\,Numerical}\\
	\includegraphics[width=0.28\textwidth]{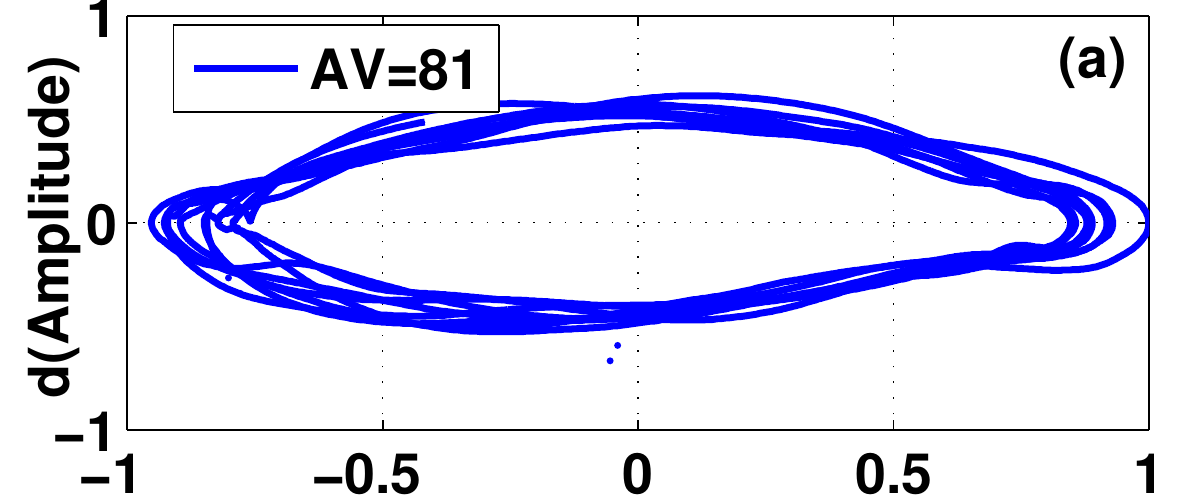}
	\includegraphics[width=0.28\textwidth]{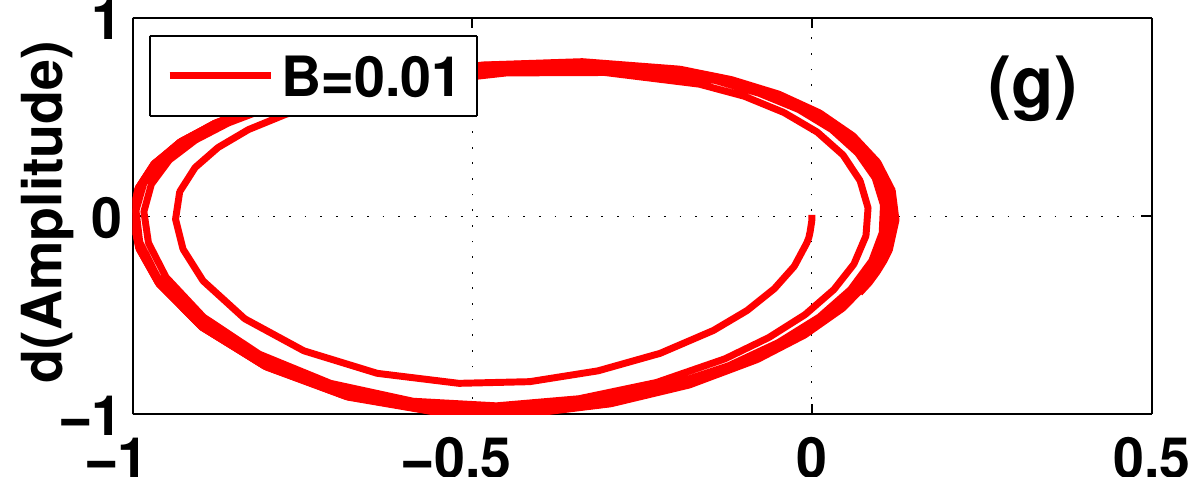}\\
	\includegraphics[width=0.28\textwidth]{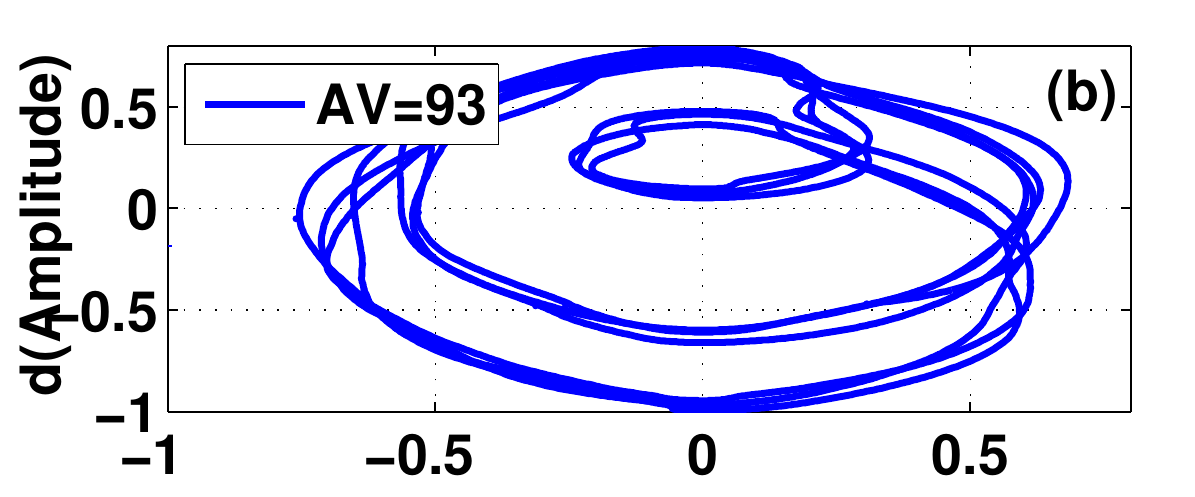}
	\includegraphics[width=0.28\textwidth]{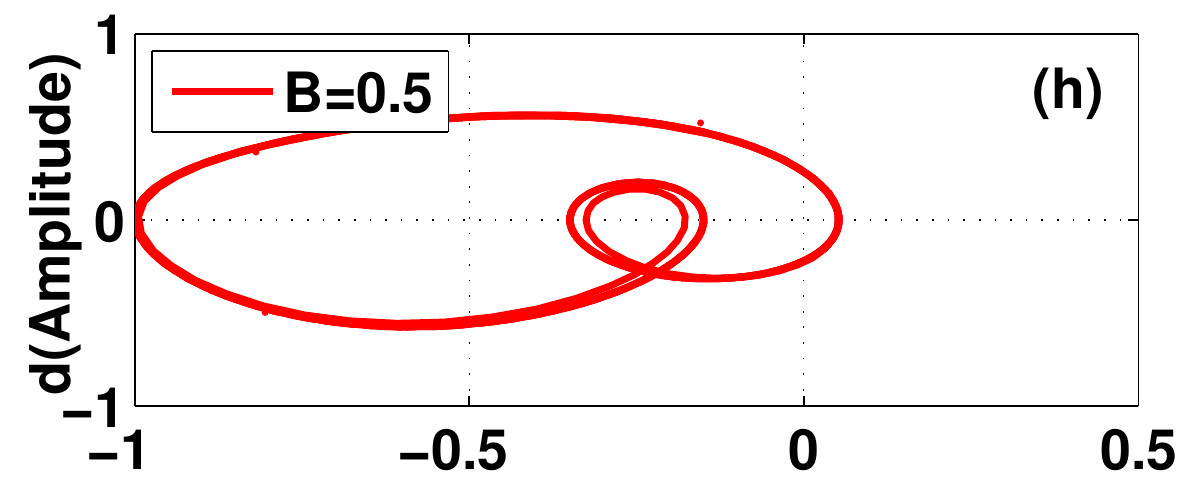}\\
	\includegraphics[width=0.28\textwidth]{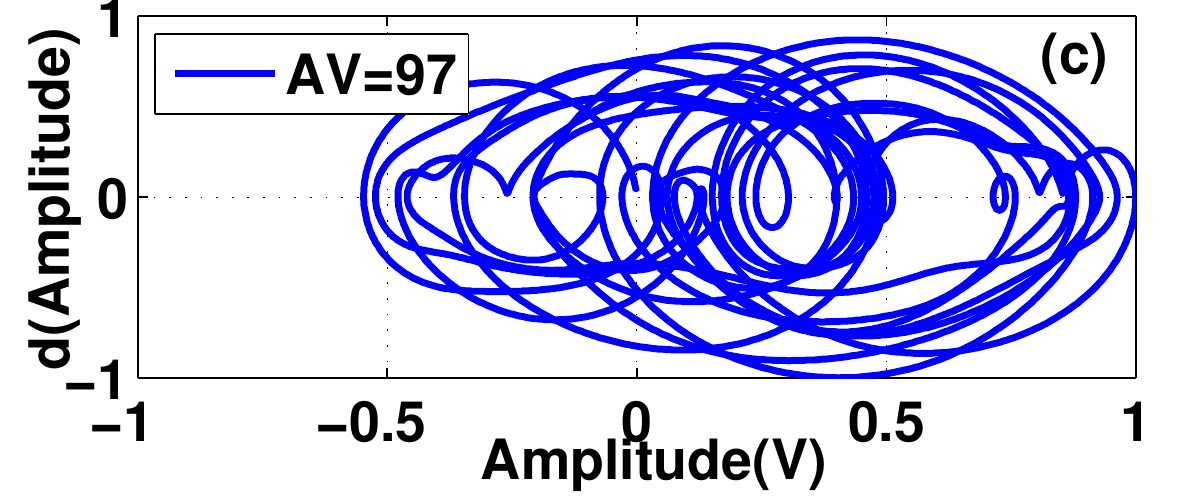}
	\includegraphics[width=0.28\textwidth]{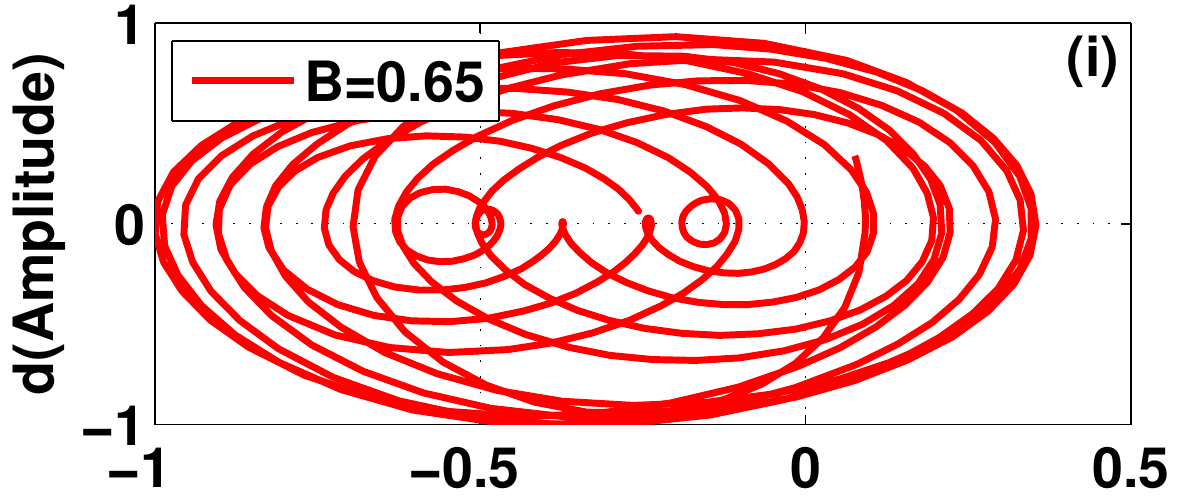}\\
	\includegraphics[width=0.28\textwidth]{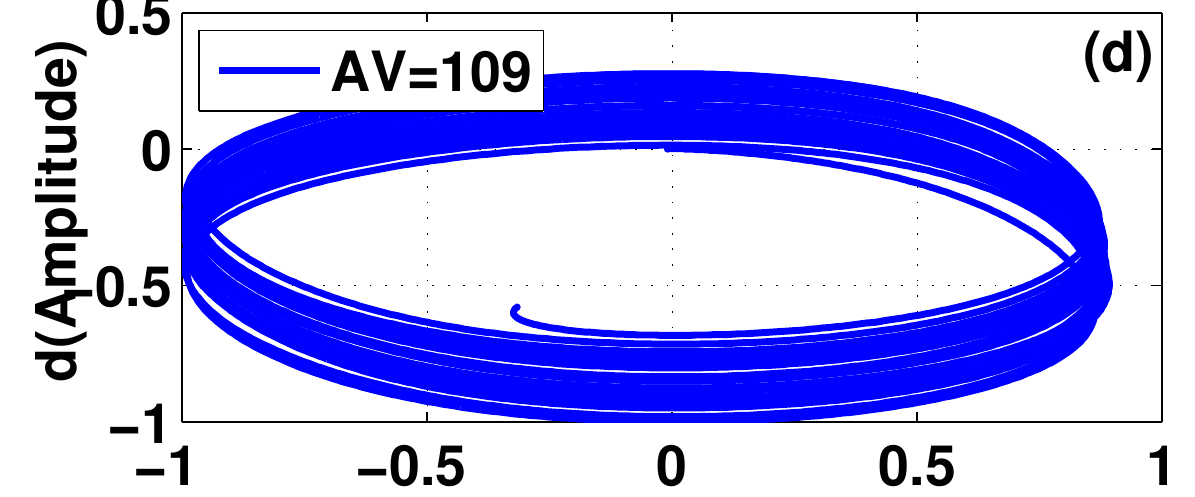}
	\includegraphics[width=0.28\textwidth]{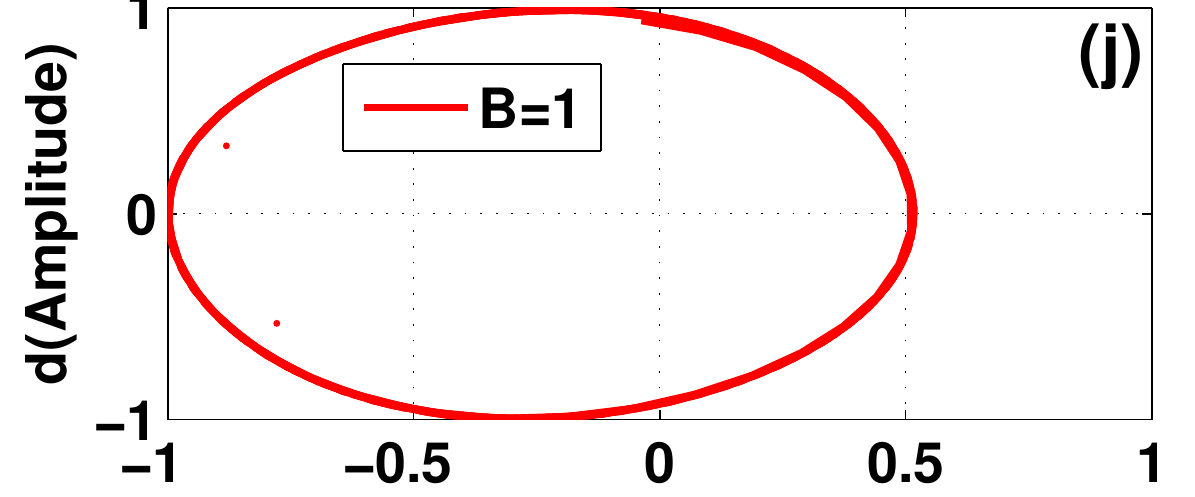}\\
	\includegraphics[width=0.28\textwidth]{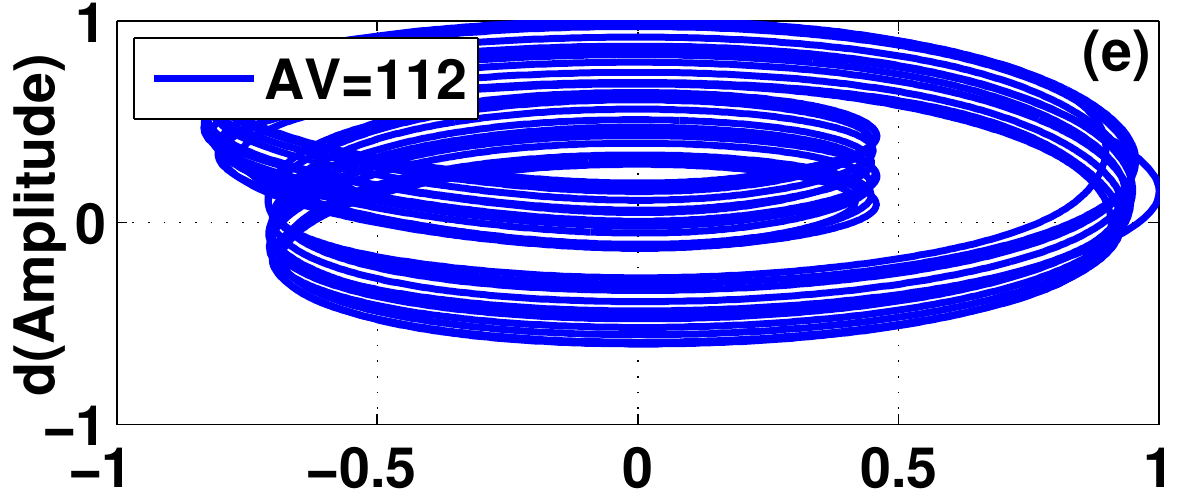}
	\includegraphics[width=0.28\textwidth]{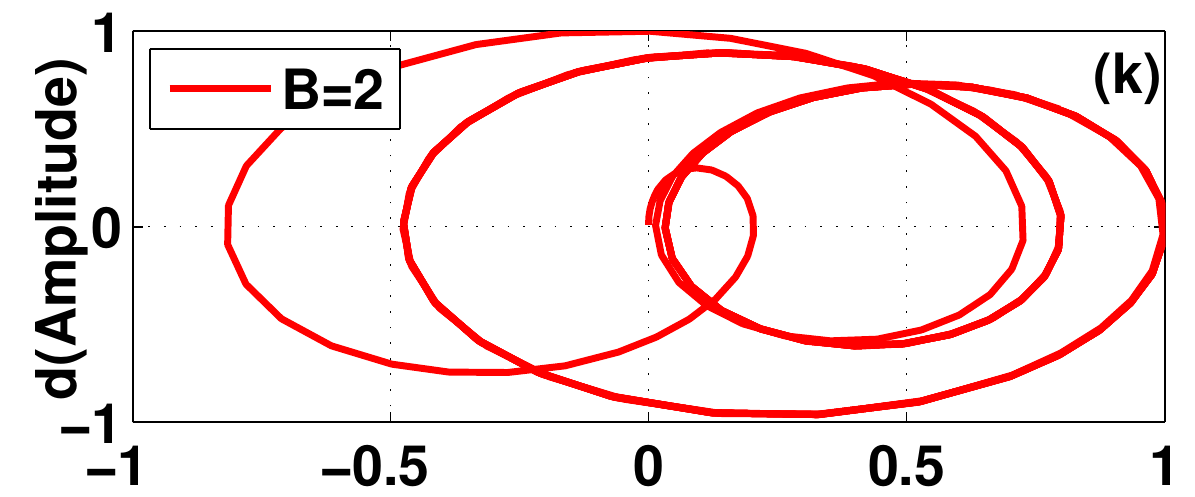}\\
	\includegraphics[width=0.28\textwidth]{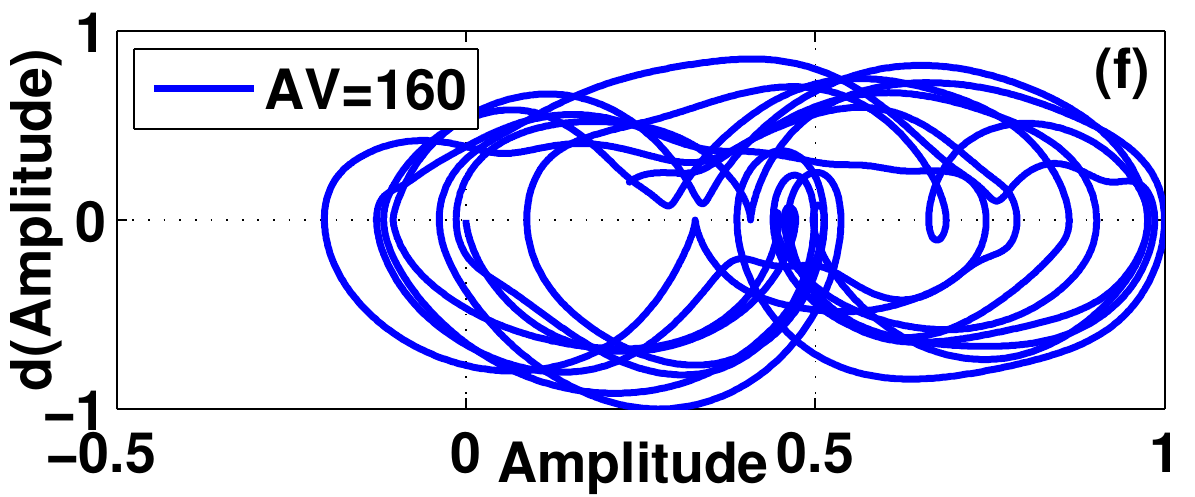}
	\includegraphics[width=0.30\textwidth]{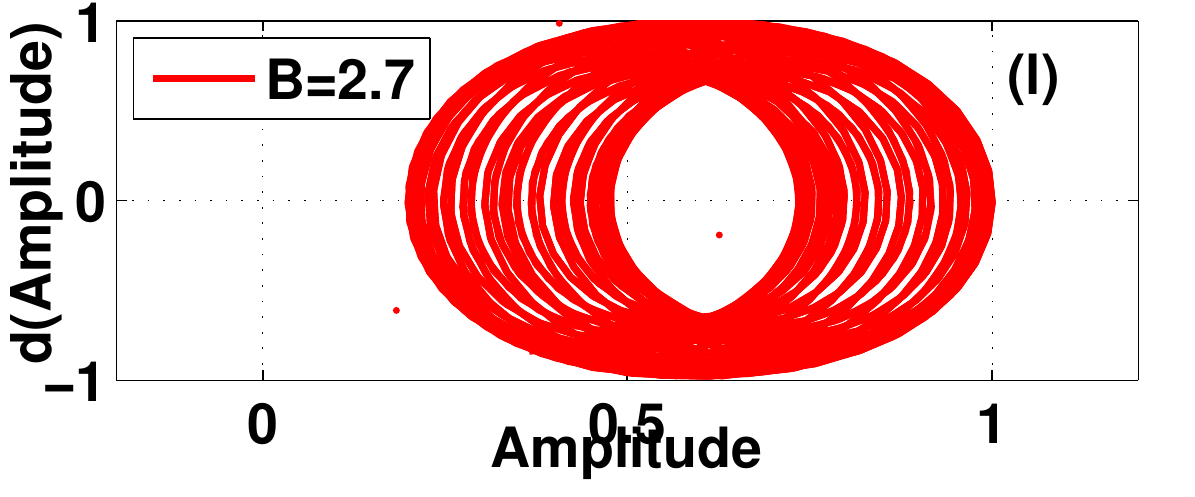}
	\caption{Phase-Space trajectory of FPF obtained by the experimentally at AV's (a) 81 V, (b)93 V, (c)97
		V ,(d)109 V (e)122 V and (f)160 V Phase-Space trajectory of FPF numerically generated by the Eq.(\ref{eqn:5}), at different parameter value of B (g) 0.01, (h) 0.5, (i) 0.65, (j) 1, (k) 2 and (l) 2.7} 
	\label{fig:3 Phase space}       
\end{figure*}

In this subsection, we present the phase space trajectories corresponding to different FPF. Phase space trajectories are a very useful tool to identify the order as well as chaotic behavior embedded in the fluctuation. The ordered nature of the fluctuation can be pin down with period-1 limit cycle motion where the trajectories after every period retrace the same path, whereas chaotic fluctuations can be recognizedwith exponentially diverging trajectories.
In Fig \ref{fig:3 Phase space}, column one is phase space plots of experimentally obtained oscillations, and column two is phase
 space plots of oscillations obtained from Eq.\ref{eqn:5}. \ref{fig:3 Phase space}(a)-\ref{fig:3 Phase space}(g) phase-space of FPF for anode voltage AV of 81 V, AV= 93 V, AV= 109 V, AV= 112 V, and AV= 160 V and Fig.\ref{fig:3 Phase space}(g)-(l) shows the
 phase-space plots of FPF obtained for different values of ($B$) 0.01, 0.5, 0.65, 1, 2 and 2.7. From these phase-space plots, it is clear that the FPFs undergo period to chaos transition through the period-doubling route. The phase-space plot also shown a significant period  doubling. Fig.\ref{fig:3 Phase space}(a) \& Fig.\ref{fig:3 Phase space}(g) exhibits the periodic nature and the Fig.\ref{fig:3 Phase space}(b) \& Fig.\ref{fig:3 Phase space}(h) shows period-doubling nature.
Fig.\ref{fig:3 Phase space}(c) \& Fig.\ref{fig:3 Phase space}(i) shows the chaotic nature of obtained time-series signals. Increasing the anode voltage leads
 the natures of FPFs to periodic to period-doubling nature and then chaos as it is indicated in Fig.\ref{fig:3 Phase space}(d) \& \ref{fig:3 Phase space}(j), \ref{fig:3 Phase space}(e) \& \ref{fig:3 Phase space}(k), and \ref{fig:3 Phase space} (f) \& \ref{fig:3 Phase space}(l). A similar trend is also observed in the numerically generated phase-space plot for various $B$ values in Eq.\ref{eqn:5} The experimental and numerically generated phase-space plots show good agreement with each other.

 \section{Fast fourier transform}
  Fourier techniques transform real-time series data in the time domain into the frequency-based domain. This technique relies on the Fourier principle that any signal comprises many sinusoids having individual frequencies. Fourier series is the constant of cosine and sine integer multiplication of fundamental frequency (harmonics) \cite{ref4.74}  . The fast Fourier transform, reducing the computation time, decreasing due to the reduction of multiplication, addition, fetching, and storing the data. For example, characterizing the power spectrum on 1000 data points, needing 1000 regular intervals, and one million multiplication to reduce this complexity on the power spectrum calculation, by introducing the windowing function to reduce multiplication \cite{ref4.74,ref4.75,ref4.76,ref4.53}. $X_{j}$, is the “welch window” after adding the window $\Omega k = 2 \Omega ck=N$
 \begin{figure}[h]
 	\centering
	\small{Experiment\,\,\,\,\,\,\,\,\,\,\,\,\,\,\,\,\,\,\,\,\,\,\,\,\,\,\,\,\,\,\,\,\,\,\,\,\,\,\,\,\,\,\,\,\,\,\,Numerical}\\
 	\includegraphics[width=0.28\textwidth]{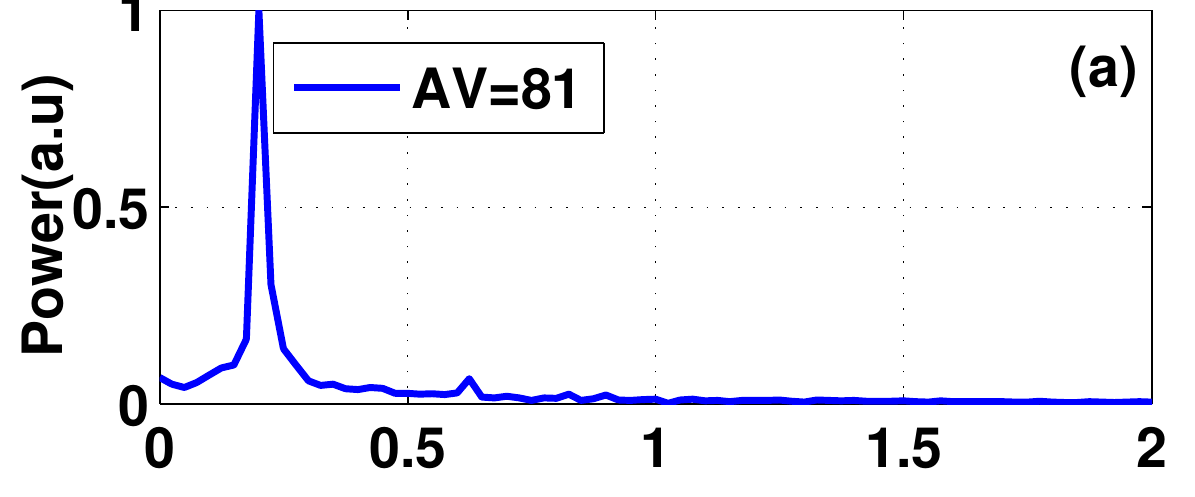}
 	\includegraphics[width=0.28\textwidth]{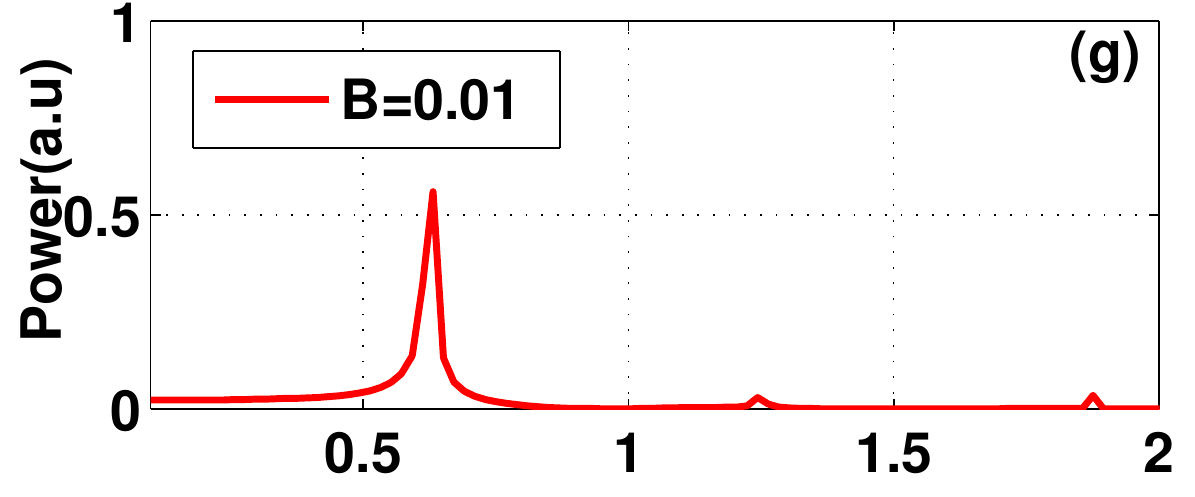}\\
 	\includegraphics[width=0.28\textwidth]{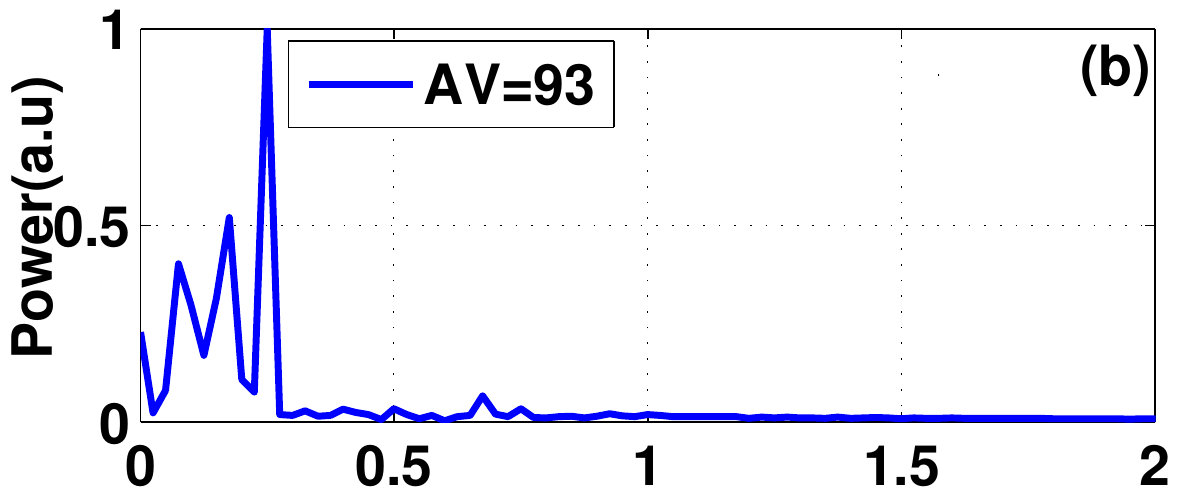}
 	\includegraphics[width=0.28\textwidth]{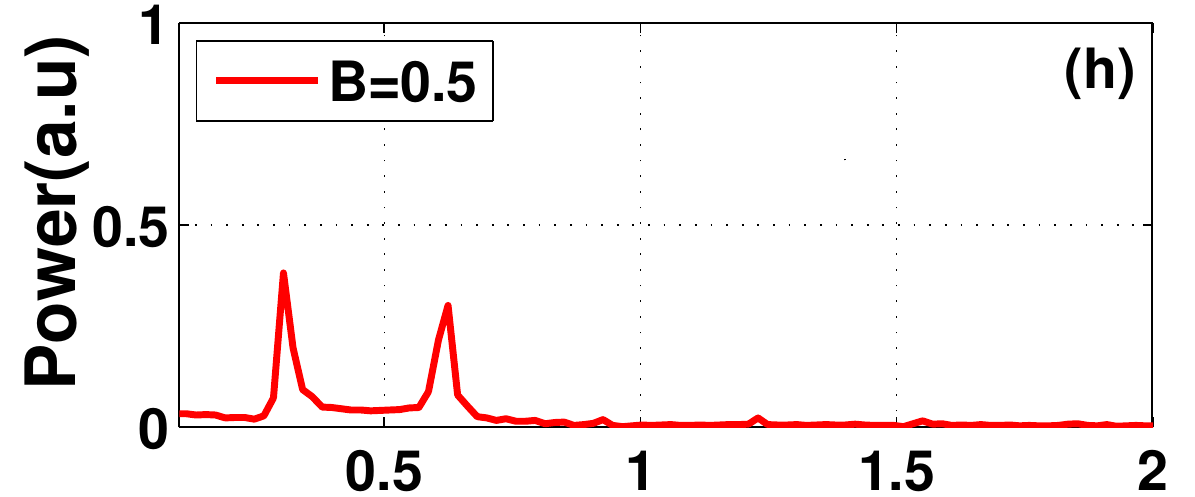}\\
 	\includegraphics[width=0.28\textwidth]{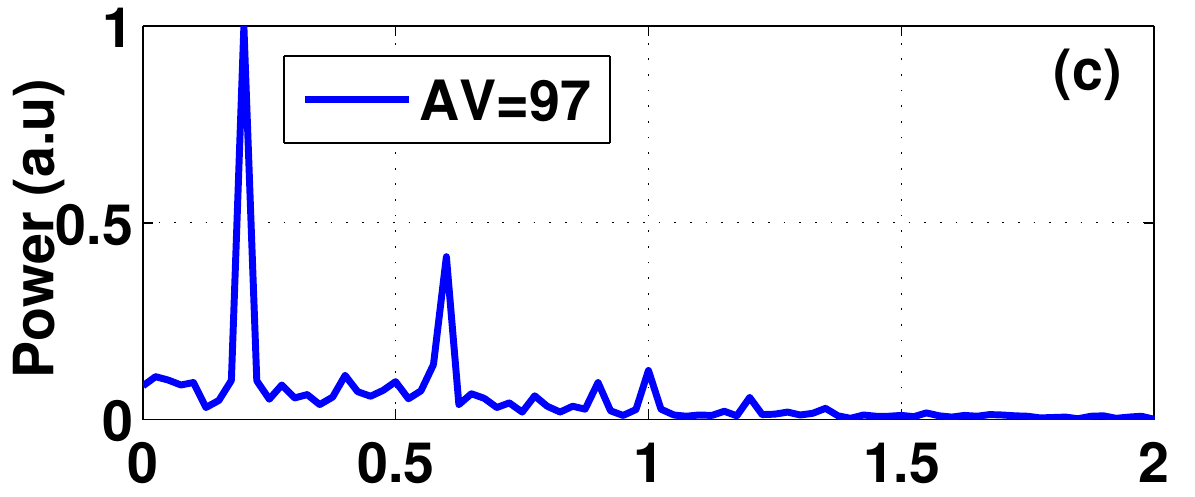}
 	\includegraphics[width=0.28\textwidth]{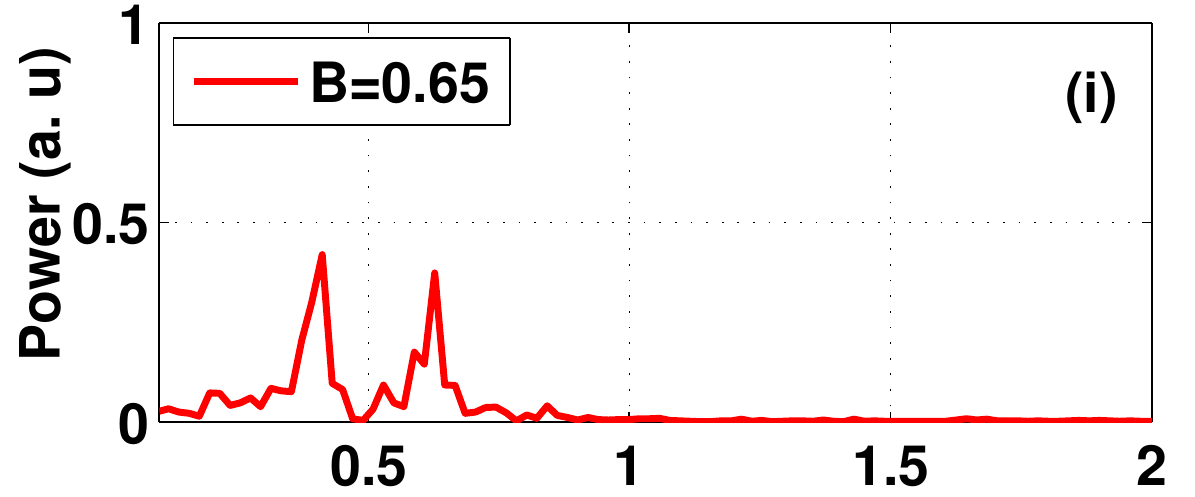}\\
 	\includegraphics[width=0.28\textwidth]{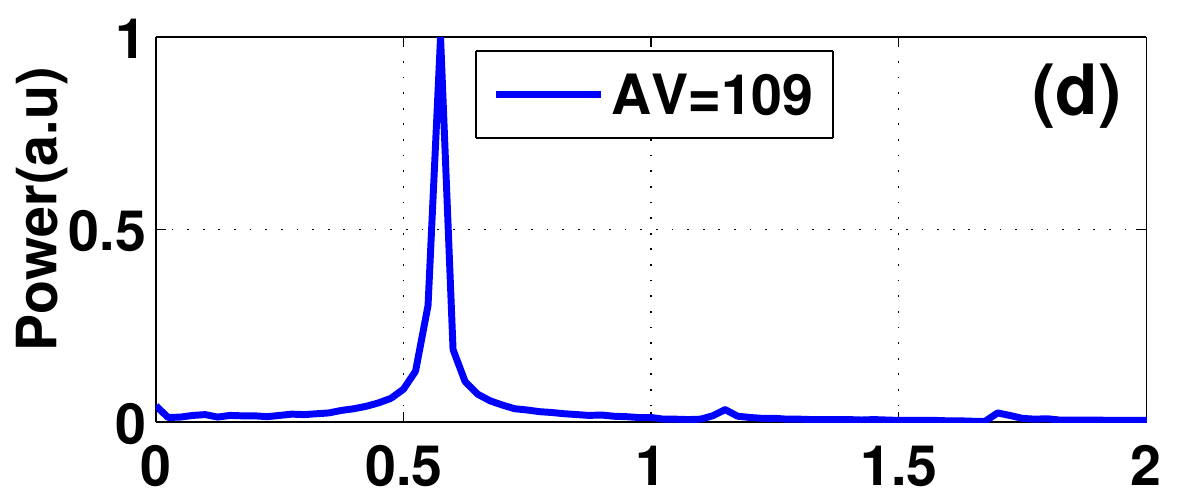}
 	\includegraphics[width=0.28\textwidth]{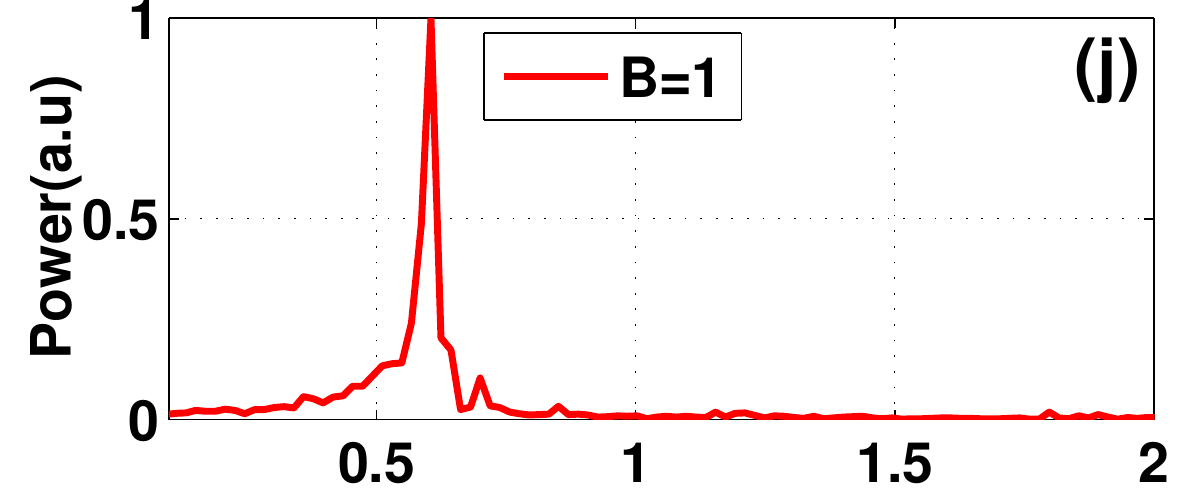}\\
 	\includegraphics[width=0.28\textwidth]{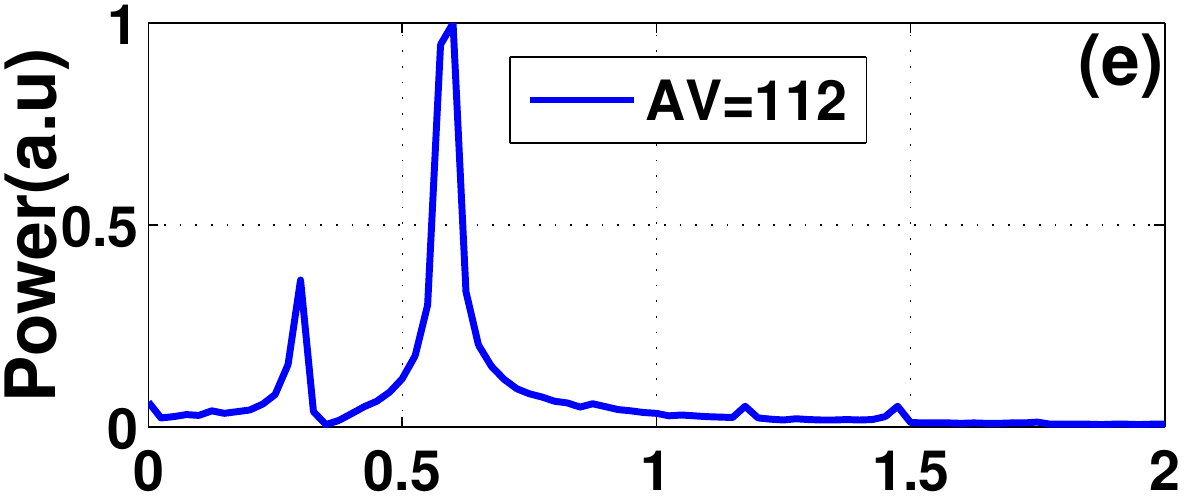}
 	\includegraphics[width=0.28\textwidth]{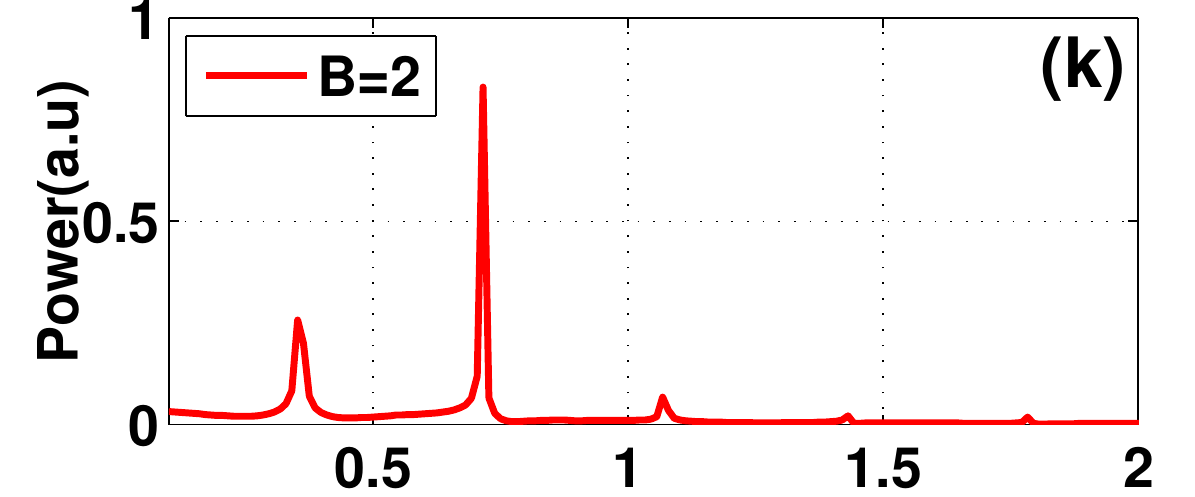}\\
 	\includegraphics[width=0.28\textwidth]{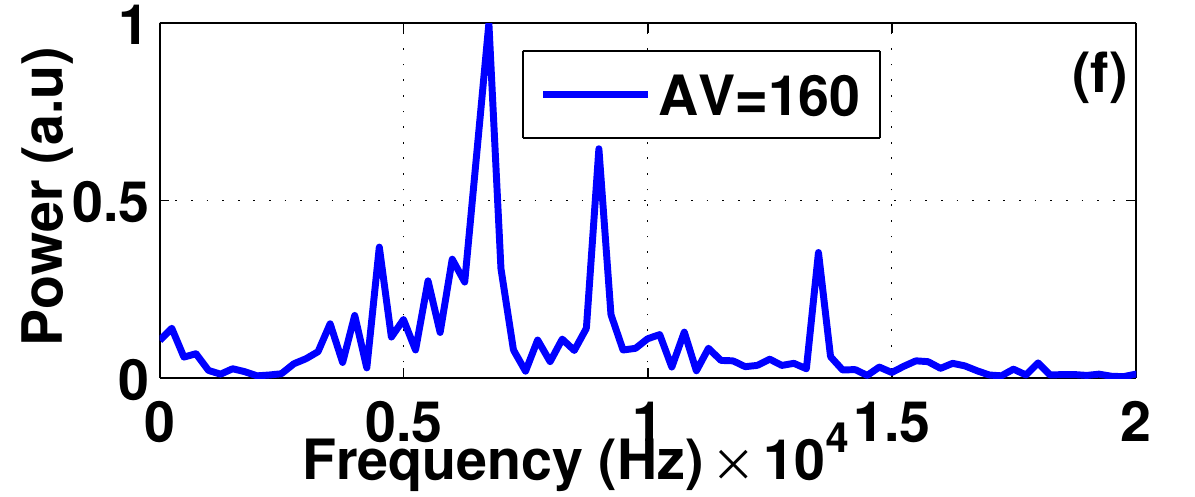}
 	\includegraphics[width=0.28\textwidth]{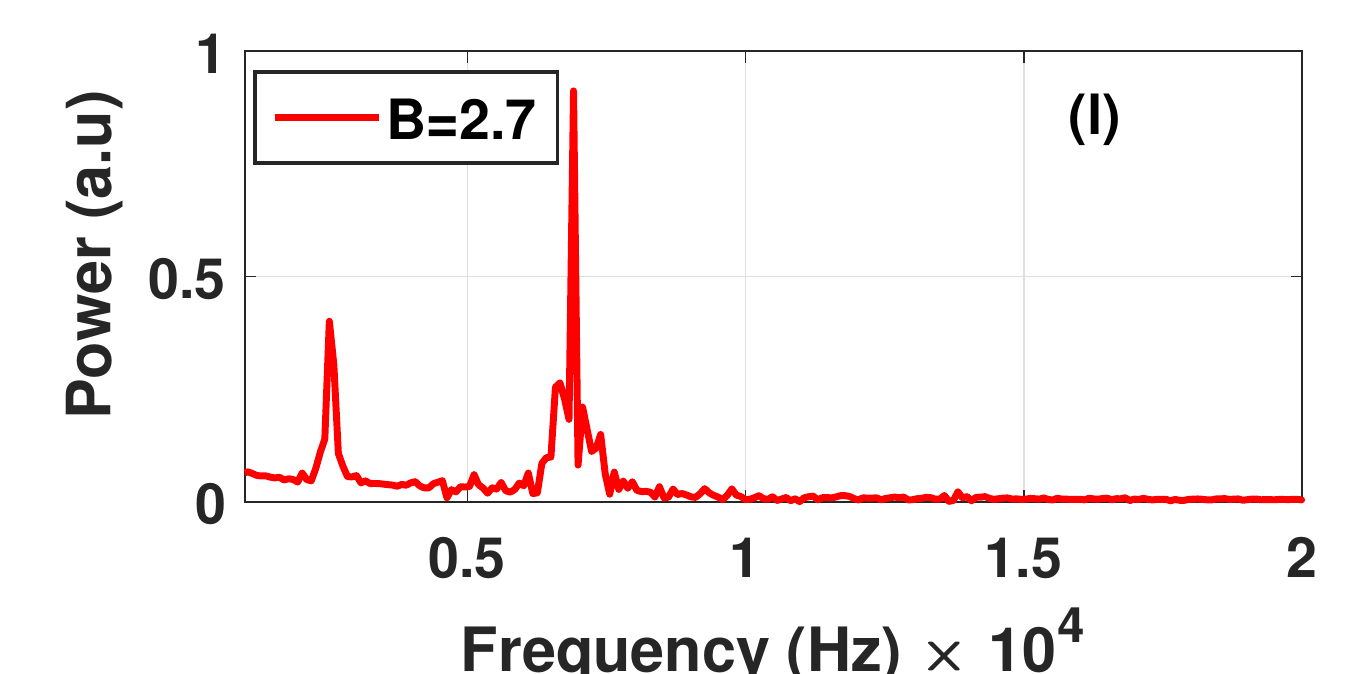}
 	\caption{In Fig. 4. The first column   plots of 2(a)-2(f) are depicted  FFT obtained by the experimentally at different AV's are:(a)81 V, (b)93 V, (c)97 V, (a)109 V, (a)112 V, (a)160 V. The second column is indicating the FFT, generated  numerically by Eq. (\ref{eqn:5}) at different amplitude forcing parameter value B's are 2(g)-2(l):(g) 0.01, (h) 0.5, (i) 0.65, (j) 1, (k) 2, (l) 2.7}
 	\label{fig:3 FFT}      
 \end{figure}

 \begin{equation}
 	C_{k}=\sum_{j=0}^{n-1}x_{i} \omega_{i} \,exp(i2\pi jk/N)\,\, ,k=0,1\dots n-1
 \end{equation}
An FFT study is carried out on the experimental time series data’s raw data and the data obtained using Eq.\ref{eqn:5}. In Fig. \ref{fig:3 FFT}, the first column indicates the frequency spectrum of experimental data obtained 
 for different applied anode voltage. The second column shows the frequency spectrum of numerically
 obtained data by Eq.\ref{eqn:5}. In Fig. \ref{fig:3 FFT}(a) and Fig. \ref{fig:3 FFT}(g), only one frequency component is present. The FPF is in periodic nature for the anode voltage 81 V and $B$ Value of 0.01. In Fig. \ref{fig:3 FFT}(b) \& Fig. \ref{fig:3 FFT}(h),one can see two different frequencies in the spectrum for the anode voltage of 93 V and the $B$ value of 0.5.
 The multiple frequencies in Fig. \ref{fig:3 FFT}(c) \& Fig. \ref{fig:3 FFT}(i) indicate the FPF for the applied anode voltage of 97 V and the $B$ of 0.65 in the Eq.\ref{eqn:5}. The anode voltage bias at 109 V, 112 V, 160 V, and $B$ values 1, 2, 2.7 exhibits periodic, followed by the period-doubling the chaotic nature. The experimentally and numerically generated FFT are found to be in good agreement with each other. 
\section{Largest lyapunov exponent}

 In linear system and the nonlinear system, the stability analysis in the time domain plays an
important role in understanding the system’s dynamics. The Lyapunov exponent is one of the standard tools
for characterizing the chaos in nonlinear dynamical systems precisely without approximation \cite{ref4.78,ref4.79,ref4.80,ref4.81}. LLE to
identify whether the time series periodic (regular) or chaotic. $x (t)$ initial time evolution at $x (0)$, appropriate state space LLE is calculated by the blow given equation.
  \begin{figure*}[h]
	\centering
	\includegraphics[width=0.40\textwidth]{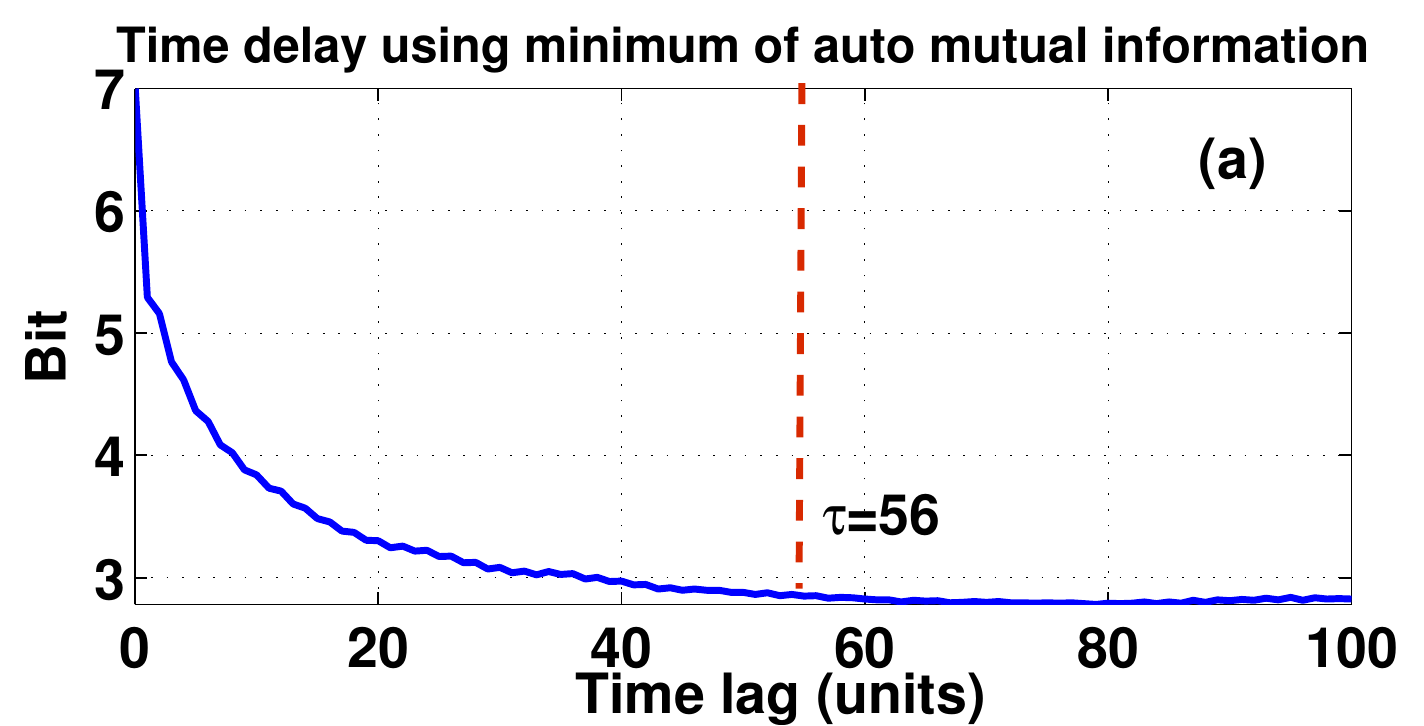}
	\includegraphics[width=0.42\textwidth]{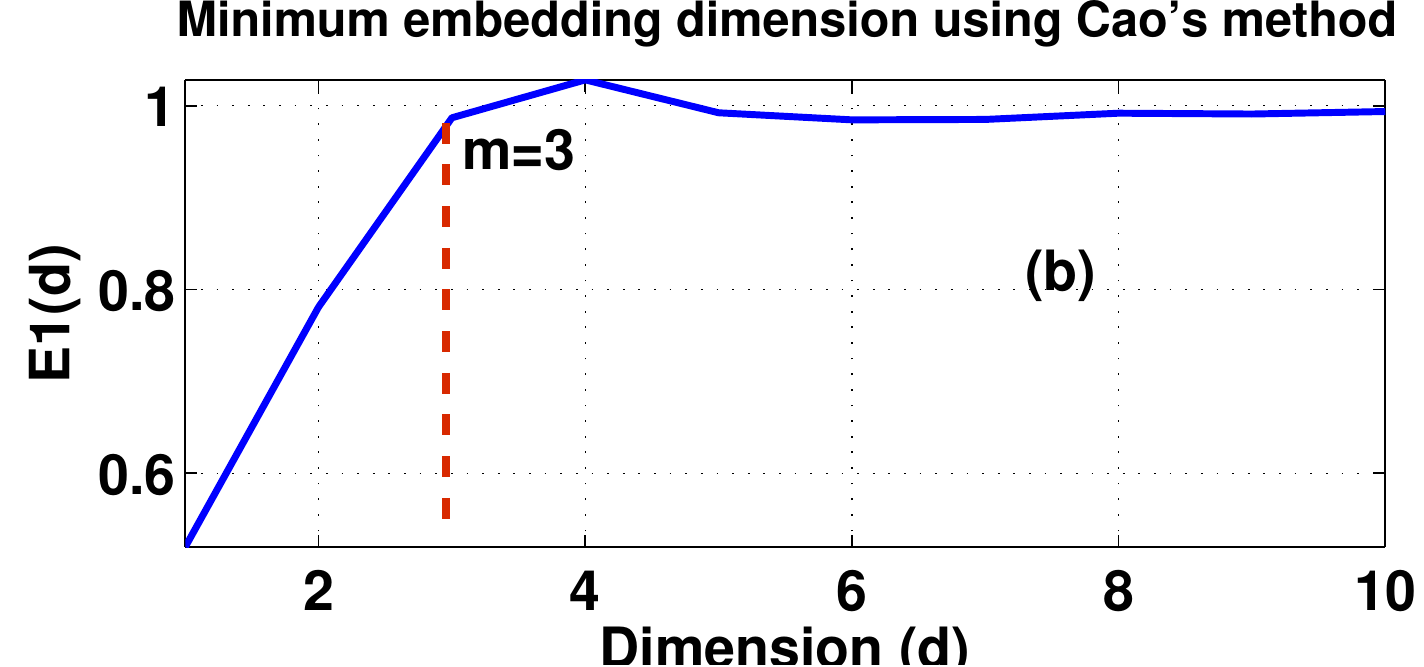}
	\caption{({\bf{a}}) is indicating the minimum time delay, calculated by the auto mutual information. ({\bf{b}}) depicted the minimum embedding dimension, calculated value of dimension as m=3.}
	\label{fig:4 TD ed}     
\end{figure*}
 \begin{figure*}[h]
 	\centering
	\includegraphics[width=0.40\textwidth]{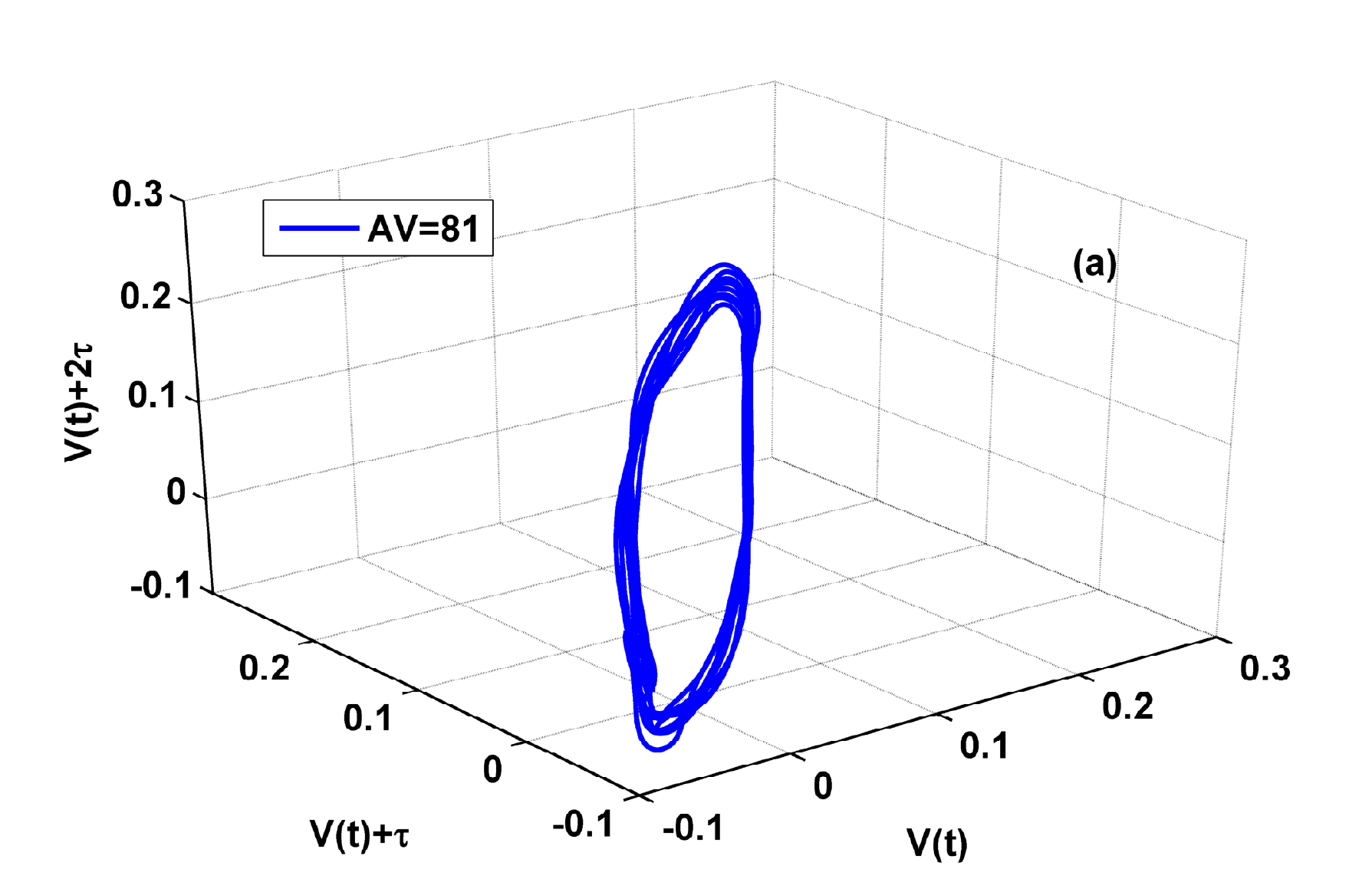}
	\includegraphics[width=0.40\textwidth]{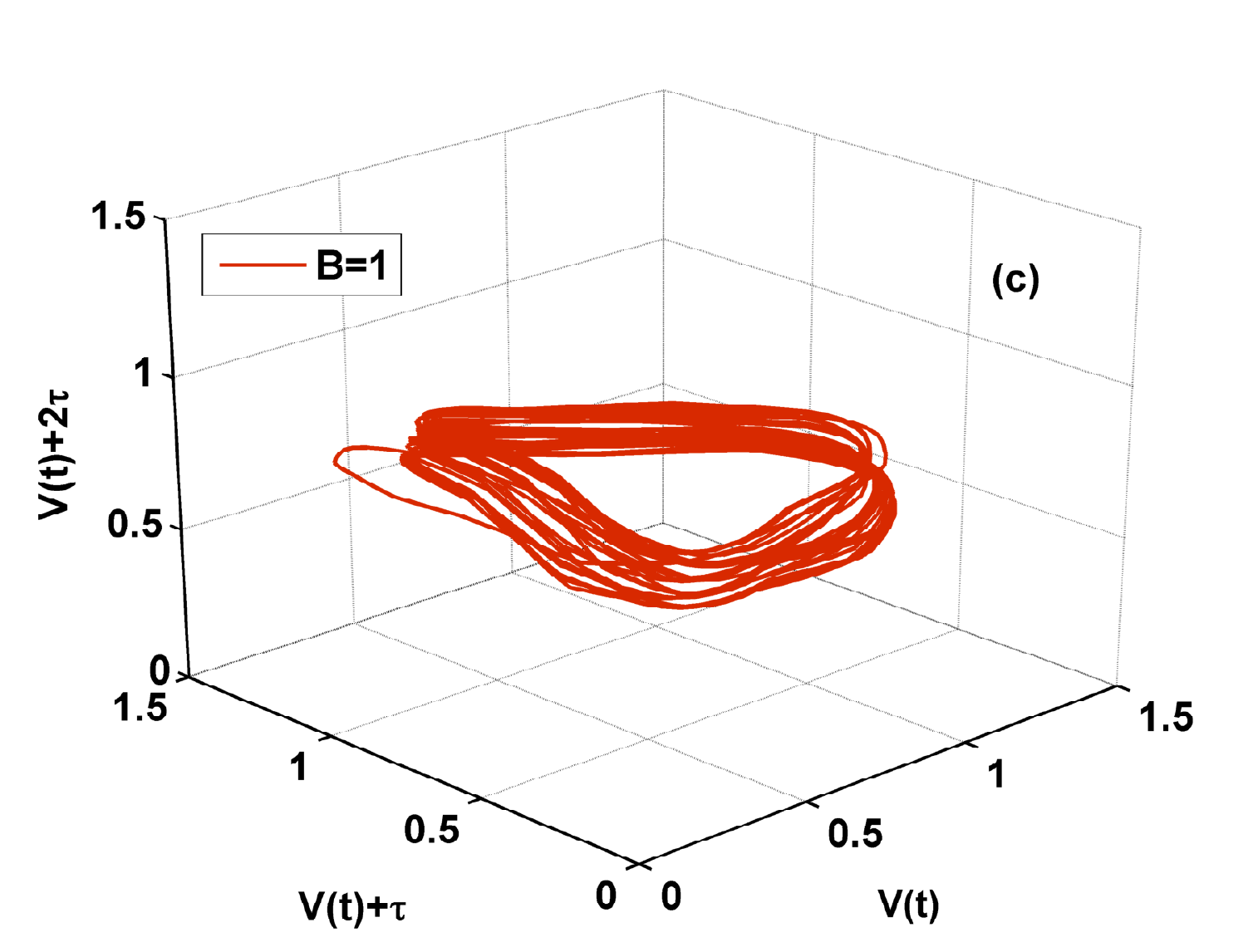}\\
	\includegraphics[width=0.40\textwidth]{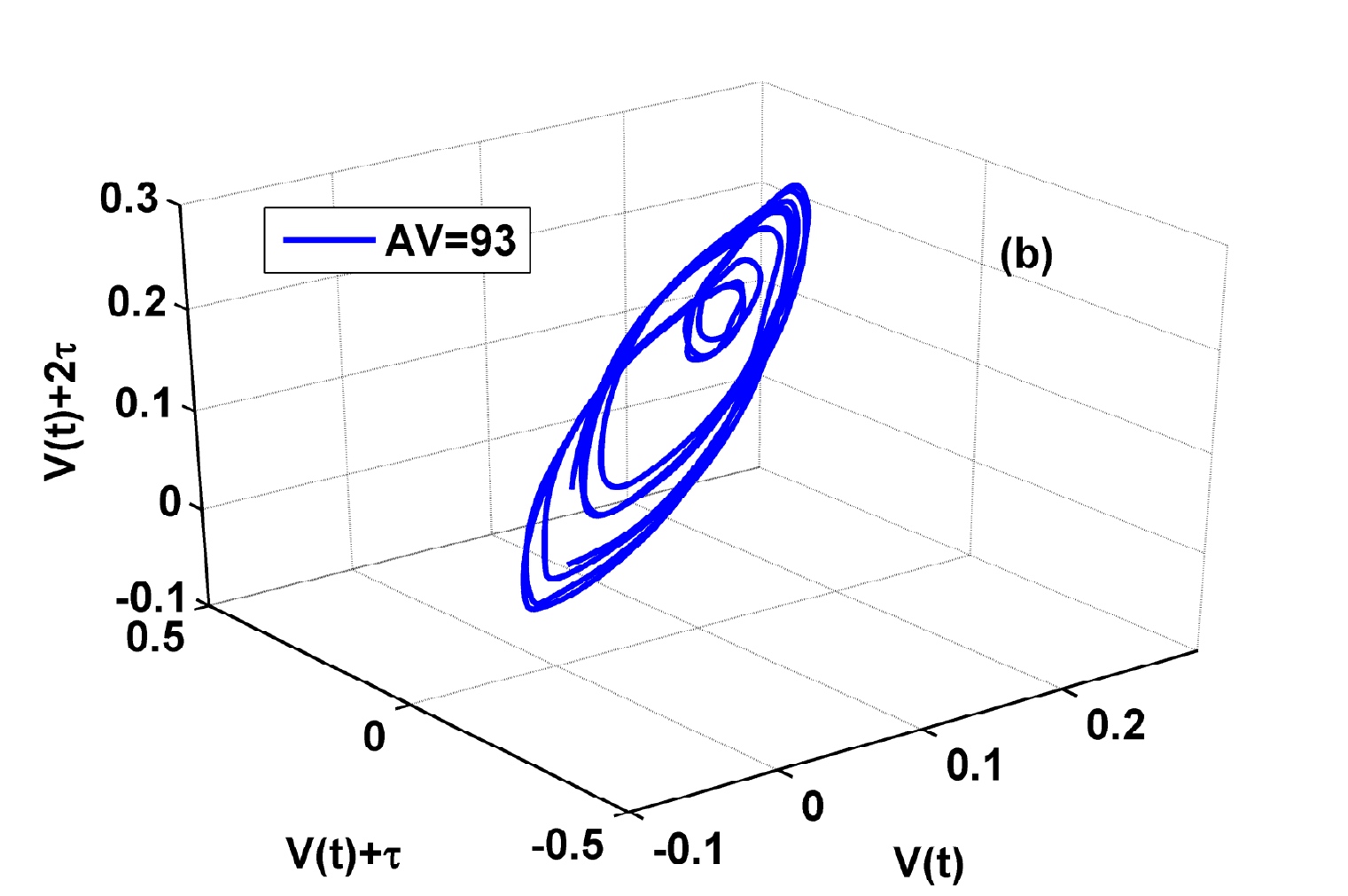}
	\includegraphics[width=0.40\textwidth]{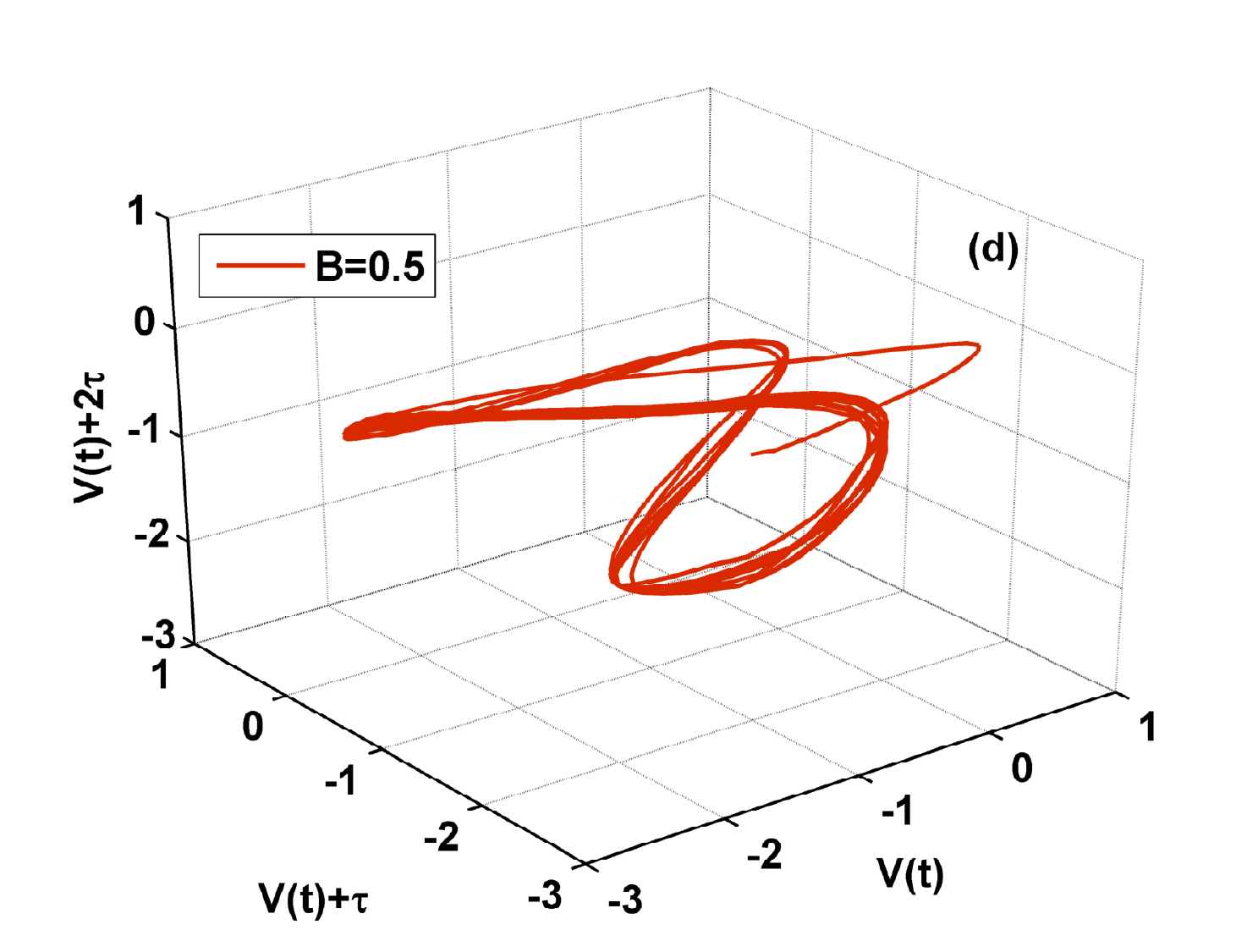}
	\caption{The vector state space reconstructed by adding the appropriate delay time of the experimental data of AV=109 and AV=112. Similarly, reconstructed the state space of numerical of $B$=0.01 and $B$=0.5 }
	\label{fig:4 SPR}     
\end{figure*}
\begin{equation}
	\lambda_{max}=\lim_{t \rightarrow \alpha} \lim_{t \rightarrow 0} \frac{l}{t}ln\left(\frac{|x(t)-x \epsilon (t)|}{\epsilon}\right)
\end{equation} 
 $S_{i} = S (t_{i}), i=1, 2,\dots N$  with $\delta t = t_{i+1}-t_{i}$ construct the vector. The state-space reconstructed, the few of numerical data and experimental data. The anode voltage is 109 V and forcing amplitude value \textbf{B} at 0.01, shown in Fig.\ref{fig:4 SPR} (a) and \ref{fig:4 SPR} (b). These figures show the multiple periods are present in the system. Similarly, \ref{fig:4 SPR} (c) and \ref{fig:4 SPR} (d) shows the multiple periods are present in the system
 \begin{equation}
S_{i}={s_{i}, \,s_{i+\tau},\,s_{i+2\tau}\dots S_{i}+(m-1)\tau }
 \end{equation}
The minimum time delay and the probable embedding dimensions are important for converting the scalar
 time series into the vector time series for the reconstruction state space and the LLE characterization in time series analysis \cite{ref4.78,ref4.79,ref4.80,ref4.81,ref4.82,ref4.83,ref4.84,ref4.85} . The space-phase of time-series data is reconstructed with the time delay
 56 ($\tau$), the obtained by using the auto mutual information is shown in Fig.\ref{fig:4 TD ed} (a). The value of $E (d)$ is saturated for $m=3$. The minimum embedding dimensions m=3 as shown in Fig.\ref{fig:4 TD ed} $(b)$
\begin{figure*}[h]
	\centering
	\includegraphics[width=0.40\textwidth]{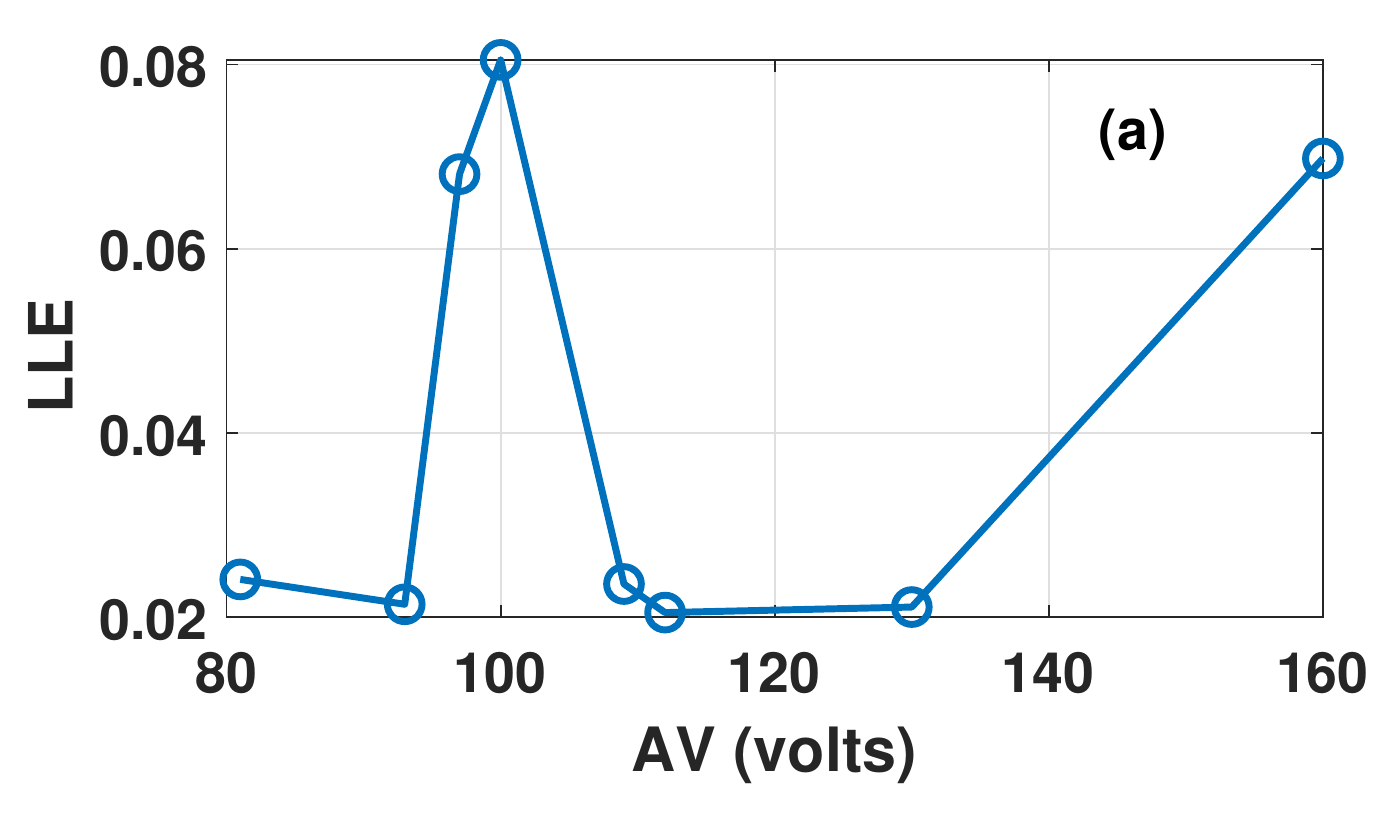}
	\includegraphics[width=0.40\textwidth]{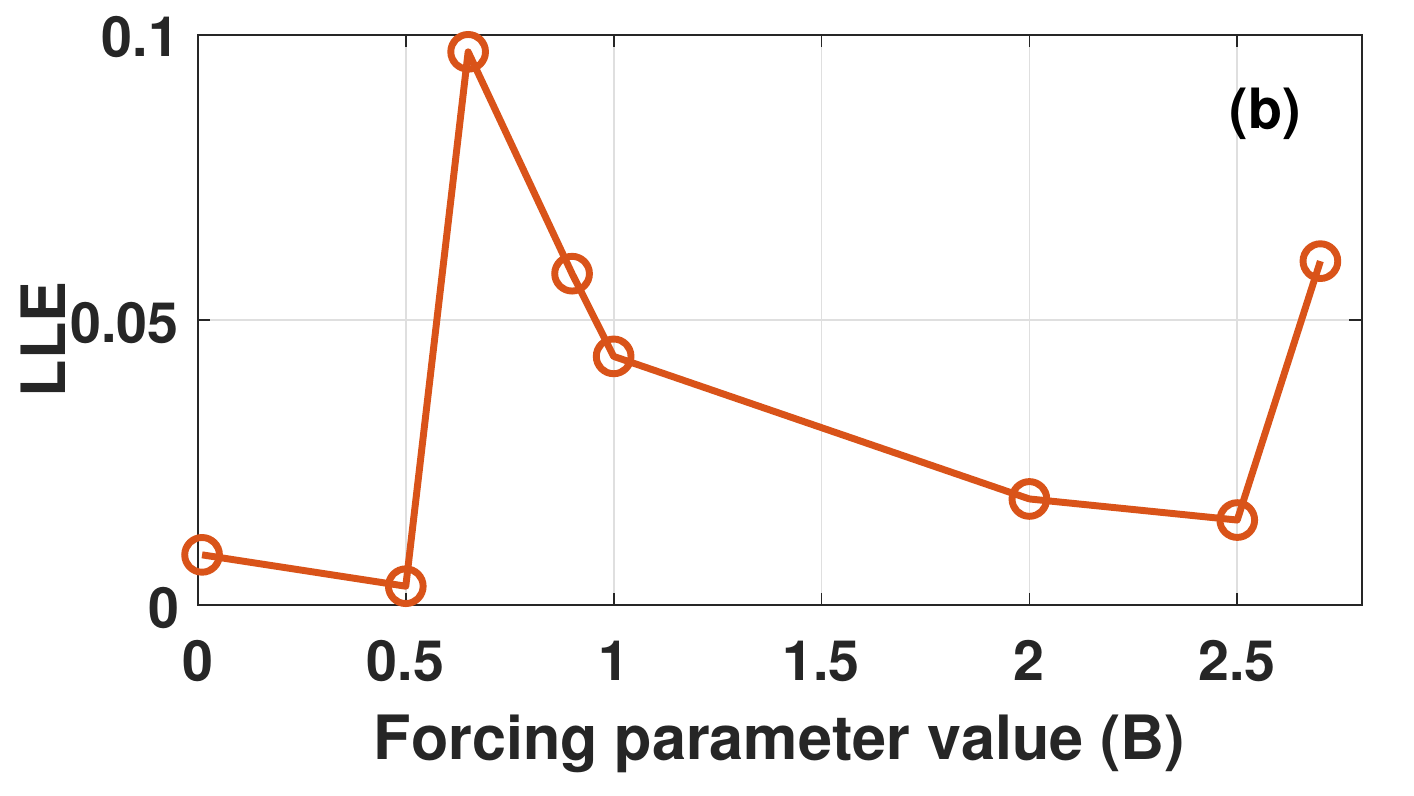}
	\caption{ Value of LLE obtained for (a) experimental and (b) numerically generated FPFs under ADL conditions is shown}
	\label{fig: LLE}      
\end{figure*}
LLE’s value gives the quantitative estimation in reconstructed phase-space divergence of convergence of
two nearby trajectories in the time series oscillations of nonlinear signals \cite{ref4.86,ref4.88,ref4.87,Kantz}  . Fig. \ref{fig: LLE} (a) shows the LLE
calculated for the FPF obtained for different applied anode voltage.  Fig.\ref{fig: LLE} (b)
depicts the LLE variation for different forcing parametervalues of $B$. LLE plot in Fig.\ref{fig: LLE} indicates the FPF at anode voltage 81 V and $B$ Value
 0.01 periodic nature. Further, increasing the anode voltage 93 V and a $B$ value of 0.5 LLE starts to increase.
 This is an indication of FPF starts to move the chaotic regime. At anode voltage 97
 V and $B$ value at 0.65. FPF shows the chaotic behavior, due to maximum values of
 LLE goes periodic to chaotic through the period-doubling. Further increasing the anode voltage and $B$ values at 109 V and 112 V, $B$ values 1 and 2, further increasing the control parameters, the multi period doubling then chaotic nature. Similarly, at anode voltage is 160 V, and $B$ value 2.7 oscillation shows the
 chaotic behavior

\section{Conclusion}
We proposed an anharmonic oscillator model with two forcing terms to study the evolution of ADL generated in
 the experimental system using two different power supplies for the cathode and the anode with respect to the grounded chamber. The experimentally observed FPFs undergo repeated periodic to chaotic to periodic transitions with monotonic increase of the applied anode voltage. The proposed DFAO model generated oscillations which demonstrated similar transitions observed in the experiment. Amplitude of the forcing term associated with cathode voltage is represented by $A$ in the model, which remained constant throughout our analysis. On the other hand, a second forcing term, denoted by $B$ is introduced into the numerical model representing the parameter to be varied, i.e., the anode bias, to model the present experimental scenario. The adopted DFAO framework successfully models the various dynamical state transition characterized by repeating period-doubling route to chaos for the forcing amplitude values that vary from the $B$=0.01, $B$=0.5, $B$=0.65, $B$=1, $B$=2, and $B$=2.7. This model is in good agreement with the dynamical characteristics experimentally observed FPF. Moreover, this numerical model mimics many characteristic features of the dc glow discharge plasma under typical ADL conditions.  Furthermore, this study may also help to stabilize the system like a double plasma device, discharge with external forcing perturbation. This work can be extended to systems with multiple forcing terms and may have significance for understanding complex systems such as biological and astrophysical DLs.

	\bibliographystyle{h-elsevier} 
	\bibliography{references}

\end{document}